\documentclass[iop,apj,numberedappendix,tighten,twocolappendix]{emulateapj}
\usepackage{epsfig}
\usepackage{epstopdf}
\usepackage{natbib}
\usepackage{appendix}
\usepackage{graphicx}
\def\ion#1#2{#1$\;${\small\rm{#2}}\relax}

\shorttitle{The $\sigma$-Discrepancy II}
\shortauthors{Rothberg et al.}

\begin{document}
\title{Unveiling The $\sigma$-Discrepancy II: \\
Revisiting the Evolution of ULIRGs \& The Origin of Quasars}
\author{Barry Rothberg\altaffilmark{1,2,3,4}, Jacqueline Fischer\altaffilmark{5}, Myriam Rodrigues\altaffilmark{6}, D. B. Sanders\altaffilmark{7}}
\altaffiltext{1}{National Research Council Postdoctoral Fellow at Naval Research Laboratory, Code 7211, 4555 Overlook Ave SW, Washington D.C. 20375, USA}
\altaffiltext{2}{Space Telescope Science Institute, 3700 San Martin Drive, Baltimore, MD 21218, USA}
\altaffiltext{3}{George Mason University, Department of Physics \& Astronomy, MS 3F3, 4400 University Drive, Fairfax, VA 22030, USA}
\altaffiltext{4}{Leibniz-Institut f\"{u}r Astrophysik Potsdam (AIP), An der Sternwarte 16, 14482, Potsdam, Germany, brothberg@aip.de, dr.barry.rothberg@gmail.com, current address}
\altaffiltext{5}{Naval Research Laboratory, Code 7211, 4555 Overlook Ave SW, Washington D.C. 20375, USA}
\altaffiltext{6}{European Southern Observatory, Alonso de Cordova 3107, Casilla 19001, Vitacura, Santiago, Chile}
\altaffiltext{7}{Institute for Astronomy, 2680 Woodlawn Drive, Honolulu, HI 96822, USA}

\begin{abstract}
\indent We present the first central velocity dispersion ($\sigma$$_{\circ}$) measured 
from the 0.85$\micron$ Calcium II Triplet (CaT) for 8 advanced (i.e. single nuclei) 
local (z $\leq$ 0.15) 
Ultraluminous Infrared Galaxies (ULIRGs).  First, these measurements are used to test the 
prediction that the ``$\sigma$-Discrepancy,'' in which the CaT $\sigma$$_{\circ}$ is 
systematically larger than the $\sigma$$_{\circ}$ obtained from the 1.6 or 2.3$\micron$ 
stellar CO band-heads, extends to ULIRG luminosities.  Next, we combine the CaT 
data with rest-frame {\it I}-band photometry obtained from  
archival {\it Hubble Space Telescope} data and the Sloan Digital Sky Survey (SDSS) 
to derive dynamical properties for the 8 ULIRGs. These are then compared to the
dynamical properties of 9,255 elliptical galaxies from the SDSS within the same 
redshift volume and  of a relatively nearby (z $<$ 0.4) sample of 53 QSO host 
galaxies.  A comparison is also made between the {\it I}-band and {\it H}-band dynamical 
properties of the ULIRGs.  We find four key results: 1) the $\sigma$-Discrepancy
extends to ULIRG luminosities; 2) at {\it I}-band ULIRGs lie on the Fundamental 
Plane (FP) in a region consistent with the most massive elliptical galaxies and not 
low-intermediate mass ellipticals as previously reported in the near-infrared;  
3) the {\it I}-band {\it M}/{\it L} of ULIRGs are consistent with an old 
stellar population, while at {\it H}-band ULIRGs appear significantly younger and less 
massive; and 4) we derive an {\it I}-band Kormendy Relation from the SDSS ellipticals 
and demonstrate that ULIRGs and QSO host galaxies are dynamically similar. 
\end{abstract}

\section{Introduction}\label{introduction}
\indent  Questions about the formation and evolution of galaxies are as challenging today 
as when the 3rd Earl of Rosse first sketched his observations of ``external nebulae''
\citep{1850PTRS...140...499}.  Referring to M51, he remarked that their 
complexity and striking beauty could hardly be the result of static processes.  Objects 
like M51 and the Antennae (NGC 4038/39) have been the focus of astronomical investigations 
since their appearance in Herschel's Catalogues of Nebulae and Clusters 
\citep{1786RSPT...76..457H}. As photographic plates replaced pencil and paper, surveys 
continued to catalog peculiar ``external nebulae'' with ever increasing speculation about 
their origins
\citep[e.g.][]{1917ApJ....46...24P,1920ApJ....51..276P,1922MNRAS..82..486P,1938MNRAS..98..613R}.
Observational work, including morphological classification and measurement of dynamical 
properties \citep[e.g.][]{1930ApJ....71..231H,1956ErNW...29..344Z}, 
the first {\it N}-body simulations \citep[e.g.][]{1941ApJ....94..385H}
and subsequent numerical simulations \citep{1965ApJ...141..768A,1972MNRAS.157..309W} 
explored the possibility that peculiar galaxies represented the
transformation of galaxies from one form into another by means of interaction.\\
\indent These earlier works all led directly to the Toomre Hypothesis 
\citep{1972ApJ...178..623T,1977egsp.conf..401T} which posits that when gas-rich spirals 
collide and merge together they form a new, more massive elliptical galaxy and that this 
process is responsible for the formation of all or most ellipticals in the Universe.  
The gravitational interaction between the two spirals rearranges the 
stellar orbits from circular to random via violent relaxation 
\citep[e.g.,][]{1967MNRAS.136..101L,1991MNRAS.253..703H}.  The process of gaseous 
dissipation funnels gas into the common gravitational center of the coalescing system, 
which triggers intense star-formation deep within molecular clouds and adds substantial 
mass to the final remnant 
\citep[e.g.,][]{1991ApJ...370L..65B,1994ApJ...437L..47M,1996ApJ...464..641M}. 
The most intense mergers are Ultraluminous Infrared Galaxies (ULIRGs), systems with 
{\it L}$_{\rm IR}$ (8-1000 $\micron$) $\geq$ 10$^{12}$ {\it L}$_{\odot}$
\citep[e.g. see][for a review]{1996ARAA..34..749S,1999Ap&SS.266..331S,1999Ap&SS.266..321J}.
The hypothesis that there is a natural evolution from ULIRG to QSO is based on the idea
that gaseous dissipation fuels more than a nuclear starburst. The accretion of 
both gas clouds and stellar remnants fuels the formation of an active galactic nucleus 
or AGN \citep[][hereafter S88]{1988ApJ...325...74S}.  This was further supported by 
the similarity between the observed bolometric ({\it L}$_{\rm Bol}$) luminosities and 
space-densities of ULIRGs and QSOs out to at least z$\sim$0.4 
\citep{1986ApJ...303L..41S,2001ApJ...555..719C}. Only after the obscuring medium in the ULIRG is cleared by 
(presumably) radiation pressure and supernovae explosions does the 
QSO become visible \citep[e.g.][]{2006ApJS..163....1H}. \\
\indent Photometric observations of spiral-spiral mergers, including ULIRGs and
their lower luminosity (10$^{11}$ {\it L}$_{\odot}$ $\leq$ {\it L}$_{\rm IR}$ 
$<$ 10$^{12}$ {\it L}$_{\odot}$) counterparts, Luminous Infrared Galaxies (LIRGs), 
demonstrated strong evidence supporting the Toomre Hypothesis.  This includes confirmation 
from optical to near-IR wavelengths that the global surface brightness (SB) profiles 
of advanced mergers follow the same de Vaucouleurs {\it r}$^{1/4}$ stellar light profile 
\citep{1953MNRAS.113..134D} that characterizes elliptical galaxies 
\citep[e.g.,][hereafter Paper I]{1982ApJ...252..455S,1990Natur.344..417W,1991A&A...245...31L,1996AJ....111..109S,1996AJ....111..655H,2002ApJS..143..315V,2004AJ....128.2098R,2006ApJ...643..707V,2010ApJ...712..318R}, 
in line with predictions from numerical simulations \citep[e.g.,][]{1988ApJ...331..699B,1992ApJ...393..484B}.
Moreover, numerical simulations show that gaseous dissipation during the merger 
will form a rotating gas disk which undergoes a strong starburst and transforms into a 
rotating stellar disk 
\citep[e.g.][]{1991ApJ...370L..65B,1994ApJ...431L...9M,1996ApJ...471..115B,1996ApJ...464..641M,2002MNRAS.333..481B,2008ApJ...679..156H}.
The starburst generates a luminosity spike at small radii, r $\leq$ 1-2 kiloparsecs (kpc), 
in the surface brightness profiles of mergers 
\citep{1994ApJ...437L..47M,2000MNRAS.312..859S}.  This {\it excess light} was first directly 
detected in the {\it K}-band SB profiles of mergers, including (U)LIRGs, and found to have 
{\it L}$_{\rm K}$ $\sim$ 10$^{9.5-10.5}$ {\it L}$_{\odot}$ 
\citep[][hereafter RJ04]{2004AJ....128.2098R}.
\cite{2008ApJ...679..156H} modeled the same sample and demonstrated that the excess light
from younger stars alone could account for 30$\%$ of the total stellar mass.  
\citep{2011AJ....141..100H} detected similar excess light at {\it H}-band using 
{\it HST} for a larger sample of (U)LIRGs spanning a wider range of merger stages.
Their Figure 14 appears to show an evolution of the excess light as a function
of the merger stage including what could be peaks at first passage and final coalescence.
These properties, taken together with the observed vast quantities 
(10$^{9-10}$ {\it M}$_{\odot}$) of cold molecular gas 
\citep[e.g.][]{1992ApJ...387L..55S,1996ApJ...457..678B,1997ApJ...478..144S,1997ApJ...484..702S,1998ApJ...507..615D,2005ApJS..158....1I,2009ApJ...692.1432G} 
and vigorous star-formation rates \citep[e.g.,][SFR]{1994ApJ...422...73P,2000ApJ...537..613A}, 
make ULIRGs prime candidates for the progenitors of giant ellipticals (gEs) and QSO hosts.\\
\indent However, a significant challenge arose for the Toomre Hypothesis and the S88 
scenario when dynamical masses ({\it M}$_{\rm Dyn}$) obtained from central velocity 
dispersions ($\sigma$$_{\circ}$) using the 1.6 or 2.3 $\micron$ CO band-heads 
(hereafter denoted as $\sigma$$_{\circ, \rm CO}$) and imaging at {\it H} (1.6 $\micron$) or 
{\it K}-band (2.2 $\micron$) implied that (U)LIRGs were the progenitors of low to 
intermediate mass ellipticals
\citep[e.g.,][]{1996ApJ...470..222S,1998ApJ...497..163S,1999MNRAS.309..585J,2001ApJ...563..527G,2002ApJ...580...73T,2006ApJ...651..835D}.
The near-infrared was used because it is less affected by the presence of dust than
optical wavelengths. \cite{2003MNRAS.340.1095D} (hereafter D03) compared 
near-IR photometry, by assuming
a fiducial $(R-K)$ transformation, with optical imaging of nearby (z $\leq$ 0.4) radio 
loud and radio quiet QSOs (RLQ and RQQ, respectively)\footnote{The dividing line between 
radio loud and quiet at 6 cm is 10$^{24}$ W Hz$^{-1}$ Sr$^{-1}$
\citep{1990MNRAS.244..207M,1992ApJ...396..487S,1995ApJ...445...62H}}. D03 concluded that 
the small half-light radii of ULIRGs precluded them as candidates for the progenitors of 
QSO host galaxies.   \cite{2006ApJ...643..707V} made a similar comparison using 
{\it Hubble Space Telescope} ({\it HST}) imaging obtained with the Near Infrared Camera and 
Multi-Object Spectrometer (NICMOS) at {\it F160W} ($\sim$ {\it H}-band) of ULIRGs and 
Palomar Green (PG) QSOs, along with QSOs from D03 
(transformed from {\it R}-band to {\it F160W}),
and relatively nearby ellipticals.  They concluded that ULIRGs were the progenitors
of 1-2 {\it L}$^{*}$ ellipticals or S0s.\\
\indent Interestingly, when the Calcium II Triplet stellar absorption lines 
($\lambda$ $\sim$ 0.85 $\micron$) were used to measure $\sigma$$_{\circ}$ 
(hereafter $\sigma$$_{\circ, \rm CaT}$) a very different picture emerged.  A comparison 
between $\sigma$$_{\circ, \rm CaT}$ and $\sigma$$_{\circ, \rm CO}$  in the same set of LIRGs 
systematically showed $\sigma$$_{\circ, \rm CaT}$ $>$ $\sigma$$_{\circ, \rm CO}$ 
\citep[][hereafter RJ06a]{2006AJ....131..185R}. Moreover, the $\sigma$$_{\circ, \rm CaT}$ 
values were consistent with ellipticals over a large mass range, including gEs.  
Recent work by \cite[][hereafter Paper I]{2009ASPC..419..273R,2010ApJ...712..318R}
has effectively explained this {\it $\sigma$-Discrepancy}.  Paper I compared 
$\sigma$$_{\circ, \rm CaT}$ with $\sigma$$_{\circ, \rm CO}$ and {\it I}-band photometry with 
{\it K}-band photometry in advanced LIRG and non-LIRG mergers as well as elliptical 
galaxies.  No $\sigma$-Discrepancy was found for elliptical galaxies, a result 
subsequently confirmed by \cite{2011MNRAS.412.2017V}.  Paper I also showed that in 
advanced mergers (Log {\it L}$_{\rm IR}$ $\leq$ 11.99), the $\sigma$-Discrepancy strongly 
correlated with Log {\it L}$_{\rm IR}$ and dust mass. Although Paper I did not include 
any ULIRGs in the sample, the results were extrapolated to brighter {\it L}$_{\rm IR}$.   
Paper I concluded that in IR-luminous mergers the near-IR observations are dominated by 
the presence of a luminous, rotating young central stellar disk (YCSD) which contains a 
population of Red Supergiant (RSG) or Asymptotic Giant Branch (AGB) stars.  Stellar disks 
have been directly detected in {\it Hubble Space Telescope} ({\it HST}) observations of 
the archetypal merger NGC 7252 \citep{1993AJ....106.1354W} and the LIRG merger NGC 34 
\citep{2007AJ....133.2132S}, with diameters ranging from 2-6 kpc. These are consistent with
the extent of the excess light at {\it K} and {\it H}-band \citep[RJ04,][]{2011AJ....141..100H};
the observed size of rotating molecular gas disks in mergers 
\citep[e.g.][]{1990AA...228L...5D,1991ApJ...368..112W,1998ApJ...507..615D} 
and numerical simulations \citep[e.g.][]{2002MNRAS.333..481B,2008ApJ...679..156H}.  
The rotating YCSD affects 
the $\sigma$$_{\circ, \rm CO}$ measured in the centers of
IR-luminous galaxies, which in turn affects the derived values of 
{\it M}$_{\rm Dyn}$.   
However, at {\it I}-band, the presence of dust, which is more centrally concentrated
due to the starburst, behaves like a coronagraph. It masks the bright YCSD so that 
$\sigma$$_{\circ, \rm CaT}$ reflects {\it only} the random motions of the old stellar 
population, probing the galaxy's true {\it M}$_{\rm Dyn}$.  Figures 13-15 in Paper I 
demonstrated that the red $(I-K)$ colors 
within the central 1.53 kpc could be best explained by many magnitudes of extinction.  
While at larger radii the $(I-K)$ colors  (Figure 13 and 14 in Paper I) were consistent 
with the average colors observed in elliptical galaxies, supporting the proposition that 
dust is centrally concentrated in IR-luminous mergers.   
Thus, when viewed at near-IR wavelengths LIRG mergers appear to have young 
stellar populations with {\it M}$_{\rm Dyn}$ $\leq$ {\it m}$^{*}$, 
where {\it m}$^{*}$ is the {\it stellar} (not dynamical) mass 
$\sim$ 3$\times$10$^{10}$ {\it M}$_{\odot}$ \citep{2003ApJ...592..819B,2003ApJS..149..289B}.
While at {\it I}-band they appear to have older stellar
populations and {\it M}$_{\rm Dyn}$ $>$ {\it m}$^{*}$.\\
\indent The results in Paper I provide strong motivation for revisiting the S88 paradigm 
of whether ULIRGs are massive enough to form gEs and host QSOs.  
This paper presents the first results for 8 ULIRGs (part of a larger survey) 
using $\sigma$$_{\circ}$ from the CaT stellar 
absorption lines in conjunction with rest-frame {\it I}-band imaging.  We probe two
important questions: 1) Does the $\sigma$-Discrepancy extend to the more luminous 
ULIRG population?  and 2)  At {\it I}-band are the dynamical properties of advanced ULIRGs 
consistent with gEs, including the host galaxies of QSOs?  \\
\indent  All data and calculations in this paper assume 
{\it H}$_{\circ}$ $=$ 75 km s$^{-1}$ Mpc$^{-1}$ and a cosmology of 
$\Omega$$_{\rm M}$ $=$ 0.3, $\Omega$$_{\rm \lambda}$ $=$ 0.7 (q$_{\circ}$ $=$ -0.55). All
photometric results are in VEGA magnitudes. In this work, ULIRGs are strictly defined as 
{\it L}$_{\rm IR}$ $\ge$ 10$^{12.0}${\it L}$_{\odot}$. LIRGs are
strictly defined as 
10$^{11.0}${\it L}$_{\odot}$ $\le$ {\it L}$_{\rm IR}$ $\le$ 10$^{11.99}${\it L}$_{\odot}$.

\section{Samples}\label{samples}
\subsection{ULIRG Sample}\label{ulirg-sample}
\indent The 8 ULIRGs analyzed in this paper were randomly selected (based on observability 
and available rest-frame {\it I}-band imaging only) from a larger, complete, volume 
limited (z $<$ 0.15) sample of 40 advanced objects taken from the Infrared Astronomical
Satellite (IRAS) 1 Jy Survey \citep{1998ApJS..119...41K} and the IRAS Revised Bright 
Galaxy Sample \citep{2003AJ....126.1607S}.  The IRAS 1 Jy survey is a complete sample 
of 118 ULIRGs down to flux levels of {\it f}$_{\nu}$ $=$ 1 Jy with Galactic latitude 
$\|${\it b}$\|$ $>$ 30$^{\circ}$, declination $\delta$ $>$ -40$^{\circ}$ 
and 0.02 $<$ z $<$ 0.27. The Revised Bright Galaxy Sample is a flux-limited survey of 
galaxies with a 60 $\micron$ flux density $>$ 5.24 Jy covering the entire sky surveyed 
by IRAS.  Late-stage ULIRGs were selected because $\sigma$$_{\circ}$ is unlikely to 
change substantially once the nuclei coalesce \citep{1999Ap&SS.266..195M}.  Based on 
numerical simulations and observations it marks the point at which the merger should 
exhibit properties in common with elliptical galaxies. HST {\it F160W} NICMOS2 
images were used to confirm the presence of a single nucleus in each system
(within the resolution limits of 59-182 pc).   Six of the eight ULIRGs were observed with 
{\it HST} using either the Advanced Camera for Surveys Wide Field Camera (ACS/WFC) or the 
Wide Field Planetary Camera 2 (WFPC2).  Photometric data for the remaining two were 
obtained from the Seventh Release of the Sloan Digital Sky Survey 
\citep[][hereafter SDSS DR7]{2000AJ....120.1579Y,2009ApJS..182..543A}.  
Optical images of the ULIRGs are shown in Figure 1 of Appendix C.  Table 1 lists the 
basic information: Names, Right Ascension (R.A.), Declination (Dec.), redshift ({\it z}), 
Log {\it L}$_{\rm IR}$, and Galactic Reddening {\it E}$(B-V)$.   {\it L}$_{\rm IR}$ is 
defined as the total flux from 8-1000 $\micron$ \citep{1996ARAA..34..749S} using the four 
{\it IRAS} passbands (12, 25, 60 and 100 $\micron$).  However, supplemental 12 or 
22$\micron$ photometry from the Wide-Field Infrared Survey Explorer was used in several 
cases where IRAS did not detect the ULIRG (see notes in Table 1).

\subsection{Comparison Samples}\label{comparison samples}
\subsubsection{SDSS {\it i}-band DR7 Elliptical Sample}\label{sdss sample}
\indent In order to assess the significance of the optically measured values of 
$\sigma$$_{\circ, \rm CaT}$ and masses of ULIRGs, a comparison sample of elliptical galaxies 
was assembled from the SDSS DR7 which offers larger spectroscopic and photometric coverage 
over earlier releases and improvements in photometric and spectroscopic measurements.  
The comparison sample was selected to be volume limited (z $\leq$ 0.15) to match the 
ULIRG sample.  A total of 9,255 elliptical galaxies were extracted from the SDSS using the 
selection criteria detailed in Appendix A. The selection criteria required that the 
elliptical galaxy must be present in both the photometric and spectroscopic databases.  

\subsubsection{Radio Loud and Radio Quiet QSOs}\label{qso sample}
\indent A relatively nearby (0.08 $<$ {\it z} $<$ 0.46) comparison sample of 28 RLQ and 
25 RQQ host galaxies was compiled from available photometry obtained with WFPC2 on 
{\it HST} and ground-based kinematic data. Only QSOs with confirmed elliptical host 
morphology were selected from the samples of 
\cite{1997ApJ...479..642B,2002ApJ...576...61H,2003MNRAS.340.1095D,2004MNRAS.355..196F,2008ApJ...678...22H}.  
The source papers all note that their samples were designed so that:  1) the RLQ and RQQ 
subsamples are matched in terms of optical luminosity; 2) {\it M}$_{\rm V}$ $<$ -23, 
representing {\it L} $\geq$ {\it L}$^{*}$ galaxies and ensuring that QSOs were selected; 
and 3) at {\it z} $\sim$ 0.4 the resolution of the WFPC2 cameras 
were sufficient to separate host from nucleus.  In all cases the authors of the source
papers performed extensive point spread function ({\it PSF}) modeling using separate 
stellar observations to properly subtract the nucleus from the host. 
Although the assembled comparison sample of QSO host galaxies is 
heterogeneous, each of the source papers have demonstrated that their samples are 
statistically representative of the local QSO host galaxy populations.  Moreover, 
the selection criteria employed by the source papers are remarkably similar with 
significant overlap.  The D03 and \cite{2004MNRAS.355..196F} (hereafter F04)
samples are sub-samples from \cite{1999MNRAS.308..377M} at 0.1 $<$ z $<$ 0.35, 
while the sample from Hamilton includes nearly all of the D03 sources, selected so that 
0.06 $<$ z $<$ 0.4.  The total integration times of all of the QSO observations were 
checked with the latest version of the WFPC2 exposure time calculator to ensure 
they were sufficiently deep to properly sample the underlying host galaxy.
The basic information for the comparison sample of RLQs and RQQs are listed in 
Columns 1-6 of Table B1 in Appendix B.    

\section{Observations and Data Reduction}\label{data reduction}
\subsection{Optical {\it F814W} Images} \label{f814w images}
\indent  The optical {\it F814W} filter was selected because the same filter was used 
for the {\it I}-band study in Paper I.  That paper demonstrated that the {\it F814W} 
filter is the best compromise between observing the old stellar population, which is used 
to probe the total {\it M}$_{\rm Dyn}$, and avoiding light produced by RSG and AGB stars.   
Here, and in Paper I the {\it F814W} filter is simply referred to as {\it I}-band.  The 
the mean differences among the ACS {\it F814W}, the WFPC2 {\it F814W}, and the Cousins 
{\it I}-band filters are less than a few hundredths of a magnitude.  
{\it F814W} images of 6 ULIRGs were obtained from the public 
{\it HST} archives (see Table 2 for more information).\\
\indent Five of the ULIRGs were observed with ACS/WFC as part of the Great Observatories 
All-Sky LIRG Survey \citep{2009PASP..121..559A}. ACS/WFC is comprised of two 
4096$\times$2048 pixel CCDs, each with a platescale of 0{\arcsec}.049 pixel$^{-1}$, 
providing a field of view (FOV) $\sim$ 202\arcsec $\times$ 202\arcsec. This FOV is large 
enough to observe each of the 5 ULIRGs completely. There is a gap of 50 pixels 
(2{\arcsec}.45) between the two CCDs.  The observations employed a two position dither
to fill the chip gap rather than the more common {\tt CR-SPLIT} (two images taken at the 
same position) which is better for cosmic ray (CR) removal but leaves the gap with no data.
As in Paper I, the ACS/WFC data were processed manually using the Image Reduction and 
Analysis Facility \footnote{IRAF is distributed by the National Optical Astronomy 
Observatory, which is operated by the Association of Universities for Research in 
Astronomy, Inc., under cooperative agreement with the National Science Foundation.} 
(IRAF) and The Space Telescope Science Data Analysis System (STSDAS), which is a software 
package designed specifically for the reduction and analysis of HST data that works with 
IRAF.  Individual exposures which have been calibrated and flat field corrected 
were obtained from the archives for the 5 ULIRGs observed with ACS/WFC.  For each ULIRG, 
the STSDAS task {\tt MULTIDRIZZLE} was used to assemble individual dithered frames into a 
final mosaic image corrected for: geometric distortions and CRs; bad pixels set to a 
value of zero; and rotated to a position angle (P.A.) of 0$^{\circ}$.   This
differs from the final drizzled image produced by the archive pipeline.  Three bad pixels 
were found in the center of IRAS 05189-2524 due to CR hits and warm pixels, and not due 
to saturation from over-exposure. The IRAF pixel editing task {\tt IMEDIT} was used to 
replace the zeroed pixels with the values interpolated from the surrounding pixels.  The 
central region of IRAS 12540+5708 was found to be saturated.  Both the diffraction spike 
generated from the bright core and a 12$\times$14 pixel rectangle in the center of 
IRAS 12540+5708 were saturated.  These were flagged by {\tt MULTIDRIZZLE} and set to a 
value of zero. No images with shorter exposures were available to replace the flagged 
pixels and they were ignored in the subsequent analysis.  Due to the two position 
dithering scheme used, {\tt MULTIDRIZZLE} was unable to flag and remove a large number 
of CRs, particularly within the chip gap.  Because the targets were centered in the 
ACS/WFC FOV, the chip gap runs through or close to the outer regions of the ULIRGs.  
As a result an algorithm was developed to remove CRs and is detailed in Appendix D.\\
\indent One ULIRG, IRAS F02021-2103, was observed with WFPC2, 
which is comprised of four 800$\times$800 pixel CCDs.  Three of the chips 
(WF2, WF3, and WF4) have a platescale of 0{\arcsec}.099 and the fourth (PC)  
0{\arcsec}.046.  This creates a non-symmetric FOV with a gap in coverage in the upper 
right quadrant.
The observations for IRAS F02021-2103 were centered on the WF3 chip. 
The observations were reduced using calibrated and flat field corrected WFPC2 science images 
obtained from the {\it HST} archives and processed with the STSDAS tasks {\tt WARMPIX},
which fixes hot pixels, and {\tt CRREJ}, which removes CRs and combines multiple frames
into a single image.   Geometric distortions were corrected by multiplying the 
CR cleaned image with a correction image \citep{1995PASP..107..156H}.  Finally, the image
was trimmed to remove pixels vignetted by the pyramid shaped beam-splitter mirror. 

\subsection{Near-Infrared {\it F160W} Images}\label{f160w images}
\indent The {\it F160W} filter was selected because no similarly deep {\it K}-band
data were available (as used in Paper I) and because the CO band-heads at 1.6 $\micron$
were used for many objects to measure $\sigma$$_{\circ, \rm CO}$.  
Published or archival data were used only for the 5 ULIRGs with published
values of $\sigma$$_{\circ, \rm CO}$.  All observations used NICMOS with the NIC2 camera, a 
256$\times$256 pixel HgCdTe array with 0{\arcsec}.075 pixel$^{-1}$ platescale 
(19{\arcsec}.2 FOV).  Photometric data for 3/5 ULIRGs were obtained from the literature. 
IRAS 17208-0014 and IRAS 23365+3604 were analyzed from {\it HST} archival data.  \\
\indent The NIC2 data were processed manually using IRAF and STSDAS. 
The {\it raw} (data received directly from the spacecraft without processing) files
were used rather than the archive processed data 
in order to properly account for: 1) the presence of the NIC2 
coronagraph;  2)  the presence of bias jumps between the quadrants; and 3) the presence 
of electronic ``bars'' which appear as vertical stripes.  The coronagraph shifts position
over time and the anomalies vary with time requiring the individual raw frames to be 
processed manually.  First, the STSDAS task {\tt CALNICA} was used to subtract dark 
current, correct for detector non-linearity, flatfield, convert to count rates, and 
identify and reject CRs.  Next, the STSDAS task {\tt PEDSUB} was used to 
correct for bias jumps between quadrants.   This differs from the standard archival 
pipeline reduction which uses {\tt PEDSKY} to remove both bias jumps and sky background.  
Because the objects fill most of the NIC2 array, using {\tt PEDSKY} will result in 
a non-uniform over-subtraction of the background.  In cases where electronic bars and 
other anomalies were found in individual raw frames, the STSDAS task {\tt NICPIPE} was 
used instead.  It allows the user to apply some or all of the
steps from {\tt CALNICA}.  In this case, all steps except flat-fielding, conversion to 
count-rates, and CR rejection were applied to the data.  Next the data were 
processed with {\tt BIASEQ}, which corrects for drift in the bias levels during 
the course of MultiAccum exposures. The data were then processed through {\tt NICPIPE}
again, this time applying flatfields, conversion to count-rates and cosmic-ray 
rejection.  The position of the coronagraph was determined in each individual frame 
and masked using the IRAF task {\tt IMEDIT}.  Other bad columns and hot 
pixels not removed with {\tt CALNICA} or {\tt NICPIPE} were manually identified and 
masked with the IRAF task {\tt IMEDIT}.  The frames were then processed with 
{\tt MULTIDRIZZLE} in the same manner as the ACS/WFC {\it F814W} data above, producing
geometrically corrected images rotated to a P.A. of  0$^{\circ}$.
A comparison between this method and data pre-processed through the standard 
archive pipeline showed a significant improvement in signal to noise (S/N), 
including the detection of faint tidal features which would otherwise not be visible.

\subsection{Spectroscopy}\label{spectroscopy}
\indent The optical spectra for all of the ULIRGs presented here were obtained with the 
Echellete Spectrograph and Imager \cite[ESI]{2002PASP..114..851S} in echelle mode at the 
W.M. Keck II 10m observatory.  Echelle mode employs a 20{\arcsec} long slit and 
cross-dispersed spectra with simultaneous coverage of 0.3927-1.1068$\micron$ projected 
onto a 2048$\times$4096 pixel CCD.  ESI has a fixed spectral resolution of 
11.5 km s$^{-1}$ pixel$^{-1}$. The final spectral resolution scales with slit width.  
A 1\arcsec.0 slit width (6.49 pixels) was used for 7/8 ULIRGs.  This corresponds to 
{\it R} $\sim$ 4000 or $\sim$ 75 km s$^{-1}$.  A 0{\arcsec}.5 slit width (3.24 pixels) 
was used for IRAS 12540+5708.  This corresponds to {\it R} $\sim$ 8300 or 
$\sim$ 37 km s$^{-1}$.  In this paper only spectral orders containing the CaT stellar 
absorption lines (order 6 or 7 depending on redshift) were used.  The scale along the 
spatial axis for order 6 and 7 are 0{\arcsec}.168 and 0{\arcsec}.163, respectively.  
The integration time and P.A. for each ULIRG is listed in Table 2.  
Calibrations, including internal flats and Hg-Ne, Xe, and CuAr arcs were taken at the 
beginning and end of the night.  No changes were detected between 
flats and arcs taken at the start and end of night.  ESI spectra for three ULIRGs 
(IRAS 05189-2524, IRAS F10378+1108, and IRAS 11387+4116) were first presented in 
\cite{2002ApJ...570..588R} but did not include CaT $\sigma$ measurements.  \\
\indent The data were reduced with IRAF.  The reduction of the data and spectral 
extraction method used are nearly the same as those described in Section 3.1 of Paper I 
(e.g. correction to heliocentric rest velocities, spectrophotometric correction to remove 
instrumental signature and provide approximate flux calibration, continuum normalization, 
and generation of the error spectrum).  The few differences with the methodology used in 
Paper I are described here.  Due to the redshifts of the ULIRGs the CaT lines lie at 
wavelengths within a spectral region dominated by strong night-sky emission lines.  
This was corrected using the IRAF task {\tt BACKGROUND} which measures a sky spectrum
at both edges of the slit and fits it with a polynomial (in this case a 1st or 2nd order 
Chebyshev polynomial) which is then subtracted from the spectrum.  The polynomial 
fitting and subtraction is carried out column by column.  In one case, the 
redshift of IRAS F02021-2103 placed the CaT absorption lines coincident with some
tellluric absorption lines. The IRAF task {\tt TELLURIC} was used to correct for the 
presence of the telluric absorption lines using a featureless spectrophotometric 
standard star. \\
\indent The spectra of 5/8 ULIRGs were extracted in a metric aperture of diameter 
1.53 kpc (see Table 3).  This size was selected to remain consistent with the literature 
\cite[e.g.][hereafter RJ06b, and Paper I.]{1995MNRAS.276.1341J,1995ApJ...439..623S,1998AJ....116.1606P,1999ApJS..124..127P,2006AJ....131..185R,2006AJ....132..976R}
However, in order to maximize S/N, larger diameter
apertures of size 3.77 kpc and 4.05 kpc were used to extract the spectra of 
IRAS F10378+1108 and IRAS 11387+4116, respectively.  Due to the brightness of the 
IRAS 12540+5708 nucleus, the spectrum was extracted in a 0.82 kpc diameter aperture
2.08 kpc northwest of the nucleus.  This is several times 
larger than the seeing of the observations, well within the region dominated by 
stellar continuum \citep{2004ApJ...613..781D}, and the same method used by 
\cite{2002ApJ...580...73T,2006ApJ...651..835D} to measure $\sigma$$_{\circ, \rm CO}$.  
Finally, the IRAF task {\tt CONTINUUM} was used to identify
and remove residual sky lines present from imperfect background subtraction in
IRAS 17208-0014, IRAS 19542+1110, and IRAS 23365+3604.  

\section{Data Analysis}\label{data analysis}
\subsection{Photometry}\label{photanalysis}
\subsubsection{Galactic Reddening}\label{galreddening}
\indent Photometry for all ULIRGs ({\it HST} and SDSS), comparison ellipticals and 
QSO host galaxies have been corrected for Galactic reddening using 
{\it E}$(B-V)$ values from \cite{1998ApJ...500..525S} as presented in 
NED\footnote{The NASA/IPAC Extragalactic 
Database (NED) is operated by the Jet Propulsion Laboratory, California Institute of 
Technology, under contract with the National Aeronautics and Space Administration.} 
and assume {\it R}$_{\rm V}$ $=$ 3.1 \citep{1999PASP..111...63F}.  These values were then 
scaled to the appropriate photometric filters: {\it F814W} and {\it z} for the ULIRGs; 
and the native filters listed in Table B1 for the QSO host galaxies using values from 
\citep{2011ApJ...737..103S}. Scaling factors for the {\it F675W} and {\it F791W} filters 
were computed using the York Extinction Solver \citep{2004AJ....128.2144M}.  The values
used are listed in Tables 3, 4, and B1. 

\subsubsection{Measured Global Photometric Parameters}\label{hstparameters}
\indent Photometry was performed on the {\it F814W} images from ACS/WFC and WFPC2/WF3
and the {\it F160W} images from NIC2 to measure the global photometric parameters:
the effective (or half-light) radius {\it R}$_{\rm eff}$ measured in metric units of kpc, 
the mean surface brightness {\it within} the effective radius ($<$$\mu$$>$$_{\rm eff}$),
and the total absolute magnitude ({\it M}$_{\rm I}$ or {\it M}$_{\rm F160W}$).  
The fluxes were measured in circular isophotes with fixed centers using the STSDAS task
{\tt ELLIPSE}.  The position of the galaxy centers in the {\it F814W} images were 
determined from {\it F160W} NIC2 images.  As in Paper I, 
there were several cases where the nucleus was clearly visible in the {\it F160W} 
images but obscured or partially obscured in the {\it F814W} images.
Foreground stars, bad pixels, artifacts or CRs missed by {\tt MULTIDRIZZLE} were 
masked with a bad pixel mask created using the technique described in Appendix D.
Masked items were set to a value of zero and ignored in the isophote fitting and flux
measurements.  An {\it r}$^{1/4}$ de Vaucouleurs profile
was fit to the isophotes produced by {\tt ELLIPSE} for each galaxy.  These
surface brightness profiles are plotted in Figure 2 of Appendix C.   The angular effective 
radius (in arcseconds) from the de Vaucouleurs profile was converted to {\it R}$_{\rm eff}$
using the angular diameter and co-moving distance for our preferred cosmology.  
The values of $<$$\mu$$>$$_{\rm eff}$ were derived from the {\it r}$^{1/4}$ profile
fits and were corrected for cosmological dimming \citep{1930PNAS...16..511T}. 
The total {\it M}$_{\rm I}$ or {\it M}$_{\rm F160W}$ were computed by 
extrapolating the best-fit de Vaucouleurs model beyond the measured data and using
the luminosity distance for our preferred cosmology.\\
\indent  Table 3 lists the {\it M}$_{\rm I}$, {\it R}$_{\rm eff}$ and 
$<$$\mu$$_{\rm I}$$>$$_{\rm eff}$ values for the ULIRGs in the {\it F814W} filter.  
Table 4 lists the measured values of {\it R}$_{\rm eff}$ at {\it F160W} and 
{\it M}$_{\rm F160W}$ for IRAS 17208-0014 and IRAS 23365+3604. 
{\it F160W} photometric results for IRAS F02021-2103, IRAS 05189-2524, and IRAS 12540+5708,
in Table 4 were obtained from \cite{2006ApJ...643..707V}.
For each of these ULIRGs, the sources were checked to ensure that an 
{\it r}$^{1/4}$ profile was the better fit to the surface brightness profile.
The values of {\it F160W} {\it R}$_{\rm eff}$ were converted to {\it equivalent radii}, 
which is the equivalent circular radius for measurements originally made with elliptical 
isophotes.  The equivalent radii were computed using the semi-major and semi-minor axes, 
or the semi-major axis and ellipticity \citep[e.g.][]{1999BaltA...8..535M}. 

\subsubsection{Global Photometric Parameters of ULIRGs and Elliptical Galaxies from the SDSS}\label{sdssparameters}
\indent No data reduction was performed on the actual SDSS DR7 images.
Photometric data for IRAS F10378+1108 and IRAS 11387+4116 were 
extracted from the SDSS DR7.  The Sloan {\it z}-band filter was selected over
the Sloan {\it i}-band filter based on two factors: \\
\noindent 1) At the redshifts of IRAS F10378+1108 and IRAS 11387+4116 the Sloan {\it z} 
filter overlaps the rest-frame wavelength coverage of the {\it F814W} filter;   \\
\noindent 2) Because the wavelength coverage of the Sloan {\it i}-filter is bluer than
the Cousins {\it I} and {\it F814W} filters, at these redshifts flux from strong emission 
lines of [OI], H$\alpha$, [NII], and [SII] could contribute significantly to the total
observed flux for each galaxy. At the same time, the observed wavelength
coverage of the {\it z}-band filter is now blue-ward of possible contamination from 
[SIII] at 0.9069 and 0.9531 $\micron$.\\
\indent The metric equivalent radius for {\it R}$_{\rm eff}$ was computed from the SDSS 
parameters {\tt devRad}, the half-light semi-major radius measured from a de Vaucouleurs 
fit to the galaxy light; and {\tt devAB}, the axis ratio from the de Vaucouleurs best 
fit profile.  Criteria 7 in Appendix A, that a galaxy is better represented by 
a de Vaucouleurs profile rather than an exponential profile, was verified for the two
ULIRGs.  Values of $<$$\mu$$_{\rm z}$$>$$_{\rm eff}$ were computed from 
the SDSS parameters {\tt devMag}, the total apparent magnitude measured 
from the de Vaucouleurs fit to the galaxy light convolved with a double-Gaussian fit 
to the PSF; and the angular equivalent radius {\it r}$_{\rm eff}$ 
\citep[e.g. equation 7 in][hereafter HB09]{2009MNRAS.396.1171H} and includes corrections
for cosmological dimming.  {\it M}$_{\rm z}$ was also computed from {\tt devMag} using the
luminosity distance for our preferred cosmology.\\
\indent  The {\tt devMag} fluxes were converted from AB magnitudes to VEGA magnitudes 
using the task {\tt CALCPHOT} which is a part of the SYNPHOT (synthetic photometry) 
program in STSDAS 
\citep{1988nds..conf..145H,1986HiA.....7..833K,2005...Synphot...Users...Guide}.
{\tt CALCPHOT} calculates synthetic photometry for any input spectra or blackbody curves 
using any filter transmission curve. It can be used to calculate {\it k}-corrections and 
transformations between filters.  A transformation value of 
$({\it z}_{\rm VEGA}-{\it z}_{AB})$ $=$ -0.51 was used.  Due to large variations in 
the rest-frame ultraviolet and optical spectral energy distributions (SEDs) the 
{\it k}-corrections for ULIRGs at z $\sim$ 0.1 can vary by 
$\Delta$ $\pm$ 0.01-0.09 mags \citep[e.g.][]{1998ApJ...492..116S,1999AJ....117.2152T}.  
Therefore, no {\it k}-corrections were applied to the two ULIRGs.  Although the same 
spectral regions are covered by the redshifted SDSS {\it z} filter and 
the rest-frame {\it F814W} filter, the shape of the filter transmission curves for the two
filters are very different. To assess the impact of this, we tested {\tt CALCPHOT} with 
12 spectral templates:  an elliptical, S0, Sa, and Sb galaxies \citep{1996ApJ...467...38K};  
six different starburst galaxies with variations in the amount of extinction 
\citep{1994ApJ...429..582C}, including extinction similar to those observed in ULIRGs; 
and a composite spectrum of the Seyfert 2 NGC 1068, which includes ultraviolet and 
optical lines, nebular and power-law continuum and cool stars 
\citep[][]{2005...Synphot...Users...Guide}.  {\tt CALCPHOT} yielded 
$({\it F814W}-{\it z})$ $=$ 0.19 $\pm$ 0.01 and 0.18 $\pm$ 0.01 for IRAS F10378+1108 and 
IRAS 11387+4116, respectively.  The small dispersion among the different templates is due 
to the similarity of the spectral features within the rest-frame {\it F814W} wavelength 
range.  These additional transforms were also applied to the values of {\tt devMag}.  
For the remainder of the paper the observed Sloan {\it z} values for IRAS F10378+1108 
and IRAS 11387+4116 will be referred to as {\it I}-band.  Table 3 lists the 
computed {\it M}$_{\rm I}$, {\it R}$_{\rm eff}$ and $<$$\mu$$_{\rm I}$$>$$_{\rm eff}$ 
values for IRAS F10378+1108 and IRAS 11387+4116.\\
\indent Global photometric parameters for the SDSS ellipticals were
computed in a similar fashion using the SDSS DR7 {\it i}-band values for {\tt devMag}, 
{\tt devRad}, {\tt devAB}, redshift, and extinction values. 
{\it k}-corrections and a filter transformation from SDSS {\it i}-band to 
{\it F814W} filter were computed using {\tt CALCPHOT} and an elliptical galaxy template 
spectrum \citep{1996ApJ...467...38K}.   The {\it k}-corrections ranged from 0.03-0.16 mags
with a filter transform of $({\it F814W}_{\rm VEGA}-{\it i}_{\rm ABMAG})$ $=$ -0.59.
Due to the large size of the SDSS elliptical sample, the computed photometry is not 
presented in the paper, but the selection criteria is provided in Appendix A and can be 
used to retrieve the sample from the SDSS DR7.

\subsubsection{RLQ \& RQQ Host Galaxies}\label{qso analysis}
\indent The details of the data reduction methods used for these galaxies can be found in
\cite{1997ApJ...479..642B,2002ApJ...576...61H,2003MNRAS.340.1095D,2004MNRAS.355..196F,2008ApJ...678...22H}.  
The samples published in these papers were observed with WFPC2 on {\it HST} using the 
{\it F606W}, {\it F675W}, {\it F702W}, {\it F791W}, or {\it F814W} filters.  The source 
papers provide surface brightnesses {\it at} the effective radius ($\mu$$_{\rm eff}$), not 
surface brightness {\it within} the effective radius ($<$$\mu$$_{\rm I}$$>$$_{\rm eff}$), 
and absolute and apparent magnitudes of each host galaxy and nucleus or {\it PSF}.  
In the source papers, these values were transformed from their
observed filters to either rest-frame Johnsons {\it V} or Cousins {\it R}-band magnitudes.  
To avoid adding additional uncertainties to the analysis, only the apparent magnitudes of the 
host galaxies in the original {\it HST} filters were used.  They were {\it k}-corrected 
and transformed to rest-frame {\it F814W} magnitudes using {\tt CALCPHOT} with an 
elliptical galaxy template \citep{1996ApJ...467...38K}.  An elliptical template was 
selected because all of the QSO host galaxies have confirmed elliptical galaxy morphologies
and observations of the host galaxy spectra for nearly half of the sample indicate they
are dominated by the presence of an older, quiescent stellar population at optical
wavelengths
\citep{2000MNRAS.316..204H,2000ApJ...528..201C,2001MNRAS.323..308N,2007MNRAS.378...83L,2010MNRAS.408..713W}.\\
\indent With the exception of \cite{1997ApJ...479..642B} the published values of 
{\it R}$_{\rm eff}$ for the QSO host galaxies were measured from elliptical isophotes. 
These were transformed to equivalent radii using ellipticities or semi-major and minor
axes from the source papers and Hamilton ({\it private communication}) and converted
to metric values of {\it R}$_{\rm eff}$ using our preferred cosmology.  These, along
with the transformed {\it F814W} apparent magnitudes were used to compute 
$<$$\mu$$_{\rm I}$$>$$_{\rm eff}$ in the same way as for the SDSS galaxies.
{\it M}$_{\rm I}$, {\it R}$_{\rm eff}$ and $<$$\mu$$_{\rm I}$$>$$_{\rm eff}$ are listed 
in Columns 8-10 of Table B1.

\subsection{Velocity Dispersions}\label{dispersions}
\subsubsection{Measurement of $\sigma$$_{\circ}$}\label{measure-sigma}
\indent  The details of the method used to measure $\sigma$$_{\circ}$ from the
extracted one-dimensional ESI spectra are given in RJ06a, 
RJ06b and Paper I. The IDL routine {\tt VELOCDISP} described in those papers and used here 
is based on a pixel-space direct fitting method to measure $\sigma$$_{\circ}$.  This 
method is similar to the technique described in \cite{1992MNRAS.254..389R}.  Briefly, the 
template stars are convolved with a Gauss-Hermite Polynomial, which is a modified 
Gaussian \citep{1993ApJ...407..525V}.  The 18 template stars used for the fitting 
range from G1 to M7.5 giants to K1 to M5 supergiants (see Table 4 in Paper I) and
are the same stars used in Paper I for the CaT and 2.3 $\micron$ CO spectra.  
The fitting function has five parameters: the line 
strength ($\gamma$), which measures the ratio of the equivalent width of the galaxy to 
that of the template star;  the mean recession velocity ($\upsilon$$_{\circ}$), the 
central velocity dispersion ($\sigma$$_{\circ}$, defined as $\sigma$ in a 1.53 kpc
metric aperture), the skewness ({\it h}$_{\rm 3}$), and kurtosis ({\it h}$_{\rm 4}$).  
The last two parameters characterize the departures from a 
Gaussian shape. The parameters are simultaneously fit to the data over the wavelength
range 0.8480-0.8690 $\micron$ \citep{2002AJ....124.2607B}.  Bad pixel masks were used
to mask out strong emission lines or regions of imperfect background subtraction.
Table 3 shows the best-fit results for the derived $\sigma$$_{\circ}$, 
heliocentric recessional velocity ({\it V}$_{\odot}$) and best-fit template star for 
each ULIRG. The errors in Tables 4 were calculated by {\tt VELOCDISP} using the error 
spectrum for each galaxy.  A more detailed discussion of this method of error analysis 
and a comparison with Monte Carlo simulations can be found in RJ06a, RJ06b, and Paper I.  
Figure 3 in Appendix C shows the CaT spectra for the eight ULIRGs (solid line), 
over-plotted with the best-fit convolved stellar template (gray dashed line).

\subsubsection{Aperture Corrections}\label{apcorrections}
\indent In order to reduce possible errors introduced by 
measuring kinematic properties in different aperture sizes, the values of $\sigma$$_{\circ}$ 
reported in this paper are either initially measured in a common aperture diameter of 
1.53 kpc or corrected to this aperture size (see Tables 3, 4, and B2).  This also applies
to all published velocity dispersions, including the QSO host galaxies, SDSS ellipticals,
and the CO velocity dispersions of ULIRGs.  The corrections used the scaling relation
from \cite{1995MNRAS.276.1341J}:

\begin{equation} {log\;\frac{\sigma(d)}{\sigma(d_{\circ})} = \alpha\; log\;\frac{d}{d_{\circ}}} \end{equation}

\noindent where $\alpha$ = -0.04\footnote{RF10 has a typographical error that reads 
$\alpha$ $=$ 0.04 instead of -0.04}, {\it d}$_{\circ}$ $=$ 1.53 kpc and {\it d} is the 
metric diameter of the circular aperture.  This is directly applicable to the SDSS
ellipticals measured in a 3{\arcsec} diameter circular fiber.  In the case of slit 
spectroscopy for the QSOs and ULIRGs, {\it d} is computed via:

\begin{equation} {\it d}\; \simeq\; 1.025\;\times\;2\;{\sqrt{\frac{xy}{\pi}}} \; \times \; {\it n}  \end{equation}

\noindent where x and y are slitwidth and extraction aperture, {\it n} is the metric 
scale (pc or kpc) in 1{\arcsec} for the galaxy computed from the angular diameter.  
The weighted average radius along the summed portion of the slit for each QSO was taken 
from Table 3 in \cite{2008AJ....136.1587W} and the slitwidths for the CO observations
of the ULIRGs from obtained from the relevant source papers (see Table 4).  

\section{Results}\label{results}
\indent The results here first address whether the $\sigma$-Discrepancy extends to ULIRG 
luminosities, as suggested by Paper I. We then test the predictions of the 
S88 paradigm by using a combination of rest-frame {\it I}-band photometric data and 
$\sigma$$_{\circ,CaT}$ to assess whether the {\it global} dynamical properties of the 
ULIRGs are consistent with those of massive ellipticals, including the host galaxies of 
QSOs, and whether they exhibit the same significant differences between optical and 
near-IR properties.

\subsection{Extending the $\sigma$-Discrepancy to ULIRGs}\label{sigmadiscrepancy}
\indent A key result from Paper I was a demonstration that in advanced 
LIRG mergers, $\sigma$$_{\circ, \rm CaT}$ is systematically larger than 
$\sigma$$_{\circ, \rm CO}$.  The same systematic discrepancy was not observed in 
bonafide elliptical galaxies \citep[Paper I,][]{2011MNRAS.412.2017V}.  Paper I further posited that the 
$\sigma$-Discrepancy should also extend to ULIRG luminosities.  We now test this 
hypothesis by comparing the values of $\sigma$$_{\circ, \rm CaT}$ with published values of
$\sigma$$_{\circ, \rm CO}$.  Figure 1 shows the five ULIRGs in the sample which have 
published values of $\sigma$$_{\circ, \rm CO}$ (see Table 4) compared with their 
$\sigma$$_{\circ, \rm CaT}$ presented in this paper (see Table 3).  The dotted line represents 
$\sigma$$_{\circ, \rm CaT}$ $=$ $\sigma$$_{\circ, \rm CO}$.  Taking into account the errors,
the $\sigma$$_{\circ, \rm CaT}$ values plotted for the 5 ULIRGs lie 2.5-22.9$\sigma$ away 
from the expected $\sigma$$_{\circ, \rm CO}$ values.  The relative difference between 
$\sigma$$_{\circ, \rm CaT}$ and $\sigma$$_{\circ, \rm CO}$ for each galaxy 
was characterized in Paper I by the parameter {\it $\sigma$$_{\rm Frac}$}:

\begin{equation} \sigma_{\rm Frac}  = \frac{\sigma_{\circ,CaT} - \sigma_{\circ, CO}}{\sigma_{\circ, CaT}} \end{equation}

\noindent This parameter was then compared with Log {\it L}$_{\rm IR}$ 
(Figure 4 in Paper I) for both advanced mergers and bonafide elliptical galaxies.
{\it L}$_{\rm IR}$ seems a natural starting point for comparison, given that the
degree of the $\sigma$-Discrepancy appears to be greater for LIRGs than non-LIRGs,
and non-existent in the control sample of ellipticals (Figure 1 of Paper I).
The comparison between $\sigma$$_{\rm Frac}$ and Log {\it L}$_{\rm IR}$ was quantified 
using the Pearson Correlation coefficient ({\it r}) which tests the degree of 
linear correlation between two independent data sets. The value of {\it r} ranges from 
-1 to +1 (anti-correlation to perfect positive correlation).  In 
Paper I, a comparison was made for the entire sample of advanced merger remnants 
(9.51 $<$ Log {\it L}$_{\rm IR}$ $<$ 11.96), 
yielding a value of {\it r} $=$ 0.77 (a strong correlation).  The comparison sample of 
ellipticals in Paper I showed no correlation ({\it r} $=$ 0.06).
In order to determine whether this correlation extends to ULIRG luminosities
plotted in Figure 2 are $\sigma$$_{\rm Frac}$ and Log {\it L}$_{\rm IR}$ values for
the 5 ULIRGs from this paper and all advanced mergers from Paper I.  The correlation 
in Figure 2 is {\it r} $=$ 0.75$^{\pm 0.02}$.
The errors on this correlation were computed using a ``jackknife'' resampling method 
\citep{1958..AMS...29...614} in which the computation of the Pearson Correlation for 
the sample of {\it n} pairs of data points is done using {\it n} - 1 pairs of data points. 
This allows for {\it n} computations of the Pearson Correlation to be made and a standard 
deviation to be computed to test the robustness of the correlation 
(i.e. that one point may drive an apparent correlation).\\   
\indent The algorithm {\tt FITEXY} was applied to the galaxies in Figure 2. It employs a 
double-weighted least-squares (DWLSQ) fit using a $\chi$$^{2}$ minimization method 
that accounts for errors in both variables \citep[][]{1992nrfa.book.....P,1992ApJ...397...55F}. 
The result is:

\begin{equation} \sigma_{\rm Frac}  = 0.18^{\pm 0.01}\;{\rm Log}\;L_{\rm IR}\; - \; 1.79^{\pm 0.11} \;\;\; ({\rm Log}\;L_{\rm IR} \ge 9.5)  \end{equation}

\noindent This is similar to Equation 7 in Paper I\footnote{Equation 7 in
Paper I was computed using a least-squares fit with errors in Y array.  
We have recomputed the fit using {\tt FITEXY}, which produces
$\sigma$$_{\rm Frac}$ $=$ 0.18$^{\pm 0.01}$ Log {\it L}$_{\rm IR}$ - 1.78$^{\pm 0.12}$}.
The Log {\it L}$_{\rm IR}$-$\sigma$$_{\rm Frac}$ correlation first shown in Paper I
extends to ULIRG luminosities, i.e. the relative difference between $\sigma$$_{\circ, \rm CaT}$
and $\sigma$$_{\circ, \rm CO}$ grows with larger {\it L}$_{\rm IR}$.

\subsection{The Dynamical Properties of ULIRGs}\label{dynproperties}
\subsubsection{The Fundamental Plane}\label{fundamental plane}
\indent Figure 3 shows the {\it I}-band Fundamental Plane edge on 
(diagonal solid solid line).  It is a two-dimensional plane embedded within the 
three-dimensional parameter space of $\sigma$$_{\circ}$, {\it R}$_{\rm eff}$, and 
$<$$\mu$$>$$_{\rm eff}$  \citep{1987ApJ...313...59D}. Early-type galaxies lie on the FP, 
late-type galaxies do not.  A relationship similar to the FP can be derived from the 
Virial Theorem \citep[e.g.][]{1987nngp.proc..175F,1988ASPC....4..329D,1992ApJ...399..462B}.  
The FP can be used as a diagnostic tool to probe whether the dynamical properties of a 
particular galaxy, or group of galaxies are similar to those of elliptical 
galaxies. In this section the two fundamental goals are: 1) to ascertain whether the 
advanced ULIRGs lie on or close to the FP; and 2) to compare where the ULIRGs lie
relative to the QSO host galaxies.\\
\indent The FP plotted in Figure 3 is from the orthogonal fit in Table 2 of HB09 and 
was derived from $\sim$ 50,000 early-type galaxies (z $\leq$ 0.36) using photometry from 
SDSS Data Release 4 and values of $\sigma$ from SDSS Data Release 6
\citep[see][for more details]{2009MNRAS.394.1978H}. As noted in Appendix A, we used the 
same parameters as HB09 to select the comparison sample
of ellipticals but with tighter restrictions on morphology in order to select only 
ellipticals and exclude S0 galaxies.  Thus, it is a sub-sample of HB09. \\
\indent The Sloan {\it i}-band was transformed to {\it H}$_{\circ}$ $=$ 75 and to 
the {\it HST} ACS/WFC {\it F814W} filter using the same method described in Section \ref{sdssparameters}. 
In Figure 3, and subsequently for all comparisons, ULIRGs are plotted as filled circles, 
RQQs as open flattened diamonds, RLQs as open squares, and the SDSS ellipticals are  
plotted in four groups based on their values of $\sigma$$_{\circ}$ (see Figure 3 caption).  
The first group roughly corresponds to sub-{\it m}$^{*}$ ellipticals, the second to 
$\sim$ {\it m}$^{*}$ ellipticals and the last two to progressively more massive systems.  
This adds an additional dimension of information to these plots; and in the case of 
Figure 3 separates the two parameters which comprise the x-axis, revealing a gradient 
of increasing $\sigma$$_{\circ}$ from left to right across the FP.  This gradient
reflects a previously reported steepening of the slope of the FP as a function of 
$\sigma$$_{\circ}$ \citep[e.g.][HB09]{1996MNRAS.280..167J,2003AJ....125.1866B}.  Also 
of note is an apparent thickening and slight warp or curvature of the FP at small 
{\it R}$_{\rm eff}$ and low $\sigma$$_{\circ}$ which is not due solely to observational
errors \citep[][HB09]{1996MNRAS.280..167J,2003AJ....125.1866B,2009MNRAS.392.1060N}.\\
\indent To determine if ULIRGs lie on or close to the FP, the scatter of the
residuals or {\it r.m.s.} of the ULIRG sample relative to the FP is compared with those 
of the SDSS Ellipticals.  By definition, these ellipticals lie on 
the FP because they are a sub-sample of the early-type galaxies used to construct the FP 
itself.  The SDSS ellipticals have {\it r.m.s.} $=$ 0.18 dex (in units of kpc), while the 
ULIRGs have {\it r.m.s.} $=$ 0.29 dex.  The residuals of the ULIRGs range from -0.72
to 0.30 (the SDSS ellipticals range from -0.68 to 0.40), with a mean and median 
of -0.05 and 0.04, respectively, indicating no systematic offsets for the sample.  
In general, the ULIRGs lie well 
within the scatter of the SDSS ellipticals, with four lying on the FP 
(within their errors).  We conclude that the ULIRGs lie on or close to 
the FP like bonafide ellipticals.  We note that IRAS 19542+1110 is 3.9$\sigma$ from 
the FP, making it an outlier, although there are SDSS ellipticals which are similarly
distant from the plane.\\
\indent Where do ULIRGs lie on the FP and how does
their location compare with those of gEs, including QSO host galaxies?
To quantify this, we used a variation of the Kolmogorov-Smirnov (KS) two-sided
(i.e. comparison between two empirical distributions) test which is applicable to 
two-dimensional data sets 
\citep{1983MNRAS.202..615P,1987MNRAS.225..155F,1992nrfa.book.....P}.  
The KS test itself probes the Null Hypothesis that the two 
distributions to be compared have the same distribution. It is a non-parametric test, 
meaning no assumption is made about the form of the distribution except that it 
must be continuous \citep[e.g.][]{1951JASA...46...253}. A standard rejection threshold 
of 95$\%$ (also known as the 0.05 confidence level) was selected 
\citep{1925smrw.book.Fisher.....B,2004SMSI.Fisher....R.A} for the analysis.
If the Null Hypothesis can be rejected at a greater confidence it will be stated, 
otherwise a statement of rejection or non-rejection will always refer to the 95$\%$ level.
The two-dimensional form of the KS two-sided test was designed to test the empirical 
distribution of data points on a plane and provide a goodness-of-fit statistic without 
the problems which arise from binning (i.e. $\chi$$^{2}$ test) or assumption of a 
particular shape to the distribution.  The test statistics were computed using
the methods outlined in \cite{1992nrfa.book.....P} which are in turn, based on 
modifications to the KS statistic \citep{1970JRSSB...32...115}.  These allow for 
computation of the 2D statistic beyond the limited case examples provided 
in \cite{1987MNRAS.225..155F}.  We note two important caveats for the KS tests used
here and throughout the remainder of Sections \ref{fundamental plane} and \ref{mdyn-iband}.  
First, the reported 
errors for the RLQ values of $\sigma$$_{\circ}$ listed in Table B2 are significantly 
larger ($\pm$ 17-34$\%$) than those of the RQQs, ULIRGs, or SDSS ellipticals. They are 
also significantly larger than those of any other parameters examined in this paper.   
Such large errors may affect the KS tests.  A set of 10,000 Monte Carlo simulations 
were performed in which a new value of $\sigma$$_{\circ}$ was randomly generated from 
within the range of $\sigma$$_{\circ}$ $\pm$ $\Delta$$\sigma$$_{\circ}$ for each RLQ.  
The KS test was then re-run for each Monte Carlo simulation to check for changes
in the results\footnote{The same Monte Carlo tests applied to the other parameters and
applied to the errors in $\sigma$$_{\circ}$ for the ULIRGs and SDSS ellipticals yielded 
no change in the KS 1D and 2D test results}.  As an aid to the reader, the results of all 
KS tests (1D and 2D) are summarized in Appendix C in Table C1.  The values in parentheses 
in Table C1 show the percentage for which the results reamin the same in the simulations 
compared to the actual test result.  Second, although IRAS 19542+1110 is a 3.9$\sigma$ 
outlier on the FP, excluding it from the KS tests does not change the results presented 
here and throughout Sections \ref{fundamental plane} and \ref{mdyn-iband}. \\
\indent The Null Hypothesis {\it cannot} be rejected when the ULIRGs are compared with: 
the 6 RLQs and all 8 of the QSO host galaxies (RLQs and RQQs).  The Null Hypothesis 
{\it can} be rejected when the ULIRGs are compared with the {\it entire} distribution of 
the SDSS sample.  When the comparison is restricted to the SDSS ellipticals in the two
largest bins, 165-225 km s$^{-1}$ and 225-420 km s$^{-1}$, the Null Hypothesis cannot be
rejected for either bin. \\
\indent We now compare the velocity dispersions of the samples and defer comparisons for 
Log {\it R}$_{\rm eff}$ and $<$$\mu$$_{\rm I}$$>$ to Section \ref{kormendy relation} 
where the RLQ and RQQ samples are significantly larger. A standard (1D) two-sided KS test
comparison was made for $\sigma$$_{\circ}$ between the ULIRGs 
and QSOs, and the ULIRGs and SDSS ellipticals.  It should be noted that the methodology 
for the standard two-sided KS test uses the tables for small samples or equations for 
large samples originally published in \cite{1972...Biometrika} as well as the modified 
KS test which is routinely used in programming language libraries and statistical 
software (e.g. Fortran, C, C++, IDL, Python, R) for comparisons among samples of any 
size without the need for comprehensive 
tables \citep{1970JRSSB...32...115,1992nrfa.book.....P}.  If the two methods disagree, 
it will be noted, otherwise it is assumed that both methods yield the same result.\\
\indent The Null Hypothesis can be rejected when the distributions
of $\sigma$$_{\circ}$ are compared for the ULIRGs and RLQs, but only 26$\%$ of the time.
The Null Hypothesis cannot be rejected when the ULIRGs are compared with all 8 QSO host 
galaxies (6 RLQs + 2 RQQs). The Null hypothesis can be rejected (and at the 99$\%$ level) 
when comparing the distributions of $\sigma$$_{\circ}$ for the ULIRGs and the entire SDSS 
comparison sample.\\
\indent Figure 3 and the 2D KS tests show that at {\it I}-band ULIRGs lie on the FP 
in a region where {\it M} $>>$ {\it m}$^{*}$ ellipticals are found.  This is in contrast 
to pure near-IR studies which showed ULIRGs are systematically offset from the 
FP in regions dominated by low to intermediate mass ellipticals 
({\it M} $\leq$ {\it m}$^{*}$) and therefore could not be the progenitors of QSO 
host galaxies
\citep[e.g.][]{1998ApJ...497..163S,2001ApJ...563..527G, 2002ApJ...580...73T,2006ApJ...651..835D}. 
The comparison with QSO host galaxies is less clear.  The $\sigma$$_{\circ}$ distribution
may not be consistent with that of RLQs.  The uncertainty arises from the large errors
associated with values of $\sigma$$_{\circ}$ for the RLQs.  However, one cannot rule out 
similarities between ULIRGs and the QSO host galaxy population as a whole.

\subsubsection{Dynamical Masses \& Stellar Populations at {\it I}-band}\label{mdyn-iband}
\indent  Paper I demonstrated that the {\it observed} dynamical properties of LIRGs
are different at {\it I}-band and {\it K}-band, an effect not seen in elliptical galaxies.
It showed that for a given LIRG, the presence of a central, relatively young population
of RSG and/or AGB stars dominates the {\it K}-band light.  As a result, {\it M}$_{\rm Dyn}$
measured at {\it K}-band is significantly smaller than {\it M}$_{\rm Dyn}$ measured
at {\it I}-band.  The apparent effective ages also typically younger at {\it K}-band
(see Figure 9 in Paper I).   At {\it I}-band, this population is largely obscured due to
dust, permitting the kinematics of the older, more global population to dominate the
observations.  Figure 4 is similar to Figure 9 in Paper I. It shows
{\it M}$_{\rm Dyn}$ vs {\it M}/{\it L} at {\it I}-band. The ULIRGs, RLQs, RQQs, and SDSS
ellipticals are plotted in Figure 4 (same symbols as Figure 3). {\it L} 
represents the total luminosity and the masses shown are the {\it total virial} 
{\it M}$_{\rm Dyn}$ of each galaxy:

\begin{equation} {\it M}_{Dyn} = \kappa \; {\frac{\sigma_{\circ}^2 \; R_{eff}}{G}}  \end{equation}

\noindent \citep{1958...Balt...Obs,1964ApJ...139..284F, 1972ApJ...175..627R, 1981ApJ...246..680T,1982ARA&A..20..399B,1985A&A...152..315B,1986AJ.....92...72R,1988AJ.....95.1047M,1989AA...217...35B}
where {\it R}$_{\rm eff}$ is the effective radius from the de Vaucouleurs fit, 
{\it G} is the gravitational constant, and $\kappa$ $=$6 (which takes into account 
the variations in shape, size, and inclination of spheroids; the impact of 
rotation on $\sigma$$_{\circ}$; and that $\sigma$ varies with radius).  
The values for {\it M}$_{\rm Dyn}$ are listed in Table 3 for the ULIRGs 
and Table B2 for the QSO host galaxies. The vertical dotted line represents an 
{\it m}$^{*}$ elliptical galaxy.  Overlaid are two sets of models representing
the evolution of {\it M}/{\it L} for a single stellar population (SSP).
The pair of models on the left are from \cite[][hereafter M05]{2005MNRAS.362..799M}.  
The solid vector is an SSP with solar metallicity and a Kroupa initial mass function 
(IMF).  Changing from solar to either half or twice solar metallicity causes only a 
slight shift ($<$ 0.1 dex) up or down in {\it M}/{\it L}, respectively.  The vector shown 
in light grey parallel to the Kroupa vector is an SSP with solar metallicity and a 
Salpeter IMF.  On the right are updated SSP models from 
\cite[][the models are hereafter referred to as CB07]{2003MNRAS.344.1000B}.  
The solid vector is an SSP with solar metallicity and Chabrier IMF and the light grey 
vector parallel to it is an SSP with solar metallicity and Salpeter IMF.  Using sub-solar 
(0.4 Z$_{\odot}$) or more than solar (1.5 Z$_{\odot}$) metallicity decreases or increases 
the {\it M}/{\it L} values by no more than 0.15 dex.  The M05 and CB07 models generally 
agree with each other at {\it I}-band, although the latter shows more variation/degeneracy 
in {\it M}/{\it L} at {\it t} $\sim$ 1-1.2 Gyr, while M05 shows some variation/degeneracy
at 0.2-0.4 Gyr.  These are likely related to differences in their treatment of thermally
pulsing asymptotic giant branch (TP-AGB) stars
\citep[e.g. see M05,][]{2006ApJ...652...85M,2007ASPC..374..303B,2010RSPTA.368..783B}.
Such differences have a more pronounced impact on the analysis of the stellar populations
of (U)LIRGs at near-IR wavelengths 
\citep[e.g.][and Paper I]{2009ASPC..419..273R,2010ASPC..423..351R}
The horizontal placement of both vectors are for display purposes only.   \\
\indent Figure 4 demonstrates that at {\it I}-band, all of the ULIRGs have 
{\it M}$_{\rm Dyn}$ $>$ {\it m}$^{*}$.  With the exception of IRAS 19542+1110, 
the ULIRGs lie at the upper end of the mass distribution.  As noted earlier, ULIRGs are 
known to contain massive quantities of H$_{\rm 2}$ and have prodigious 
star-formation rates.  Various methods for estimating the SFR 
\citep[e.g.][]{1998ARA&A..36..189K,2001ApJ...554..803Y} indicate the 8 ULIRGs
plotted in Figure 4 have SFRs $\sim$ 100-500 {\it M}$_{\odot}$ yr$^{-1}$. 
Yet, when the {\it M}/{\it L}$_{\rm I}$ values are compared with the SSPs, the stellar
populations appear to be old and evolved for 7/8 ULIRGs.  A 2D KS test comparison of the
distribution of the ULIRGs with the galaxies in the {\it I}-band
{\it M}$_{\rm Dyn}$-{\it M}/{\it L} plane indicates that the Null Hypothesis 
can be rejected when the ULIRGs are compared with the QSO host galaxies 
(either the RLQs alone or all 8 QSOs).  However, the large RLQ $\sigma$$_{\circ}$ errors 
weaken this result significantly, especially when the entire QSO sample is considered 
(see Table C1 for details).  In addition, the Null Hypothesis can be rejected 
for the {\it entire} comparison sample of SDSS ellipticals, although the Null Hypothesis 
cannot be rejected when the ULIRGs are compared with the SDSS ellipticals in the 
165-225 km s$^{-1}$ and 225-420 km s$^{-1}$ bins. .\\
\indent It should be noted that Figure 4 presents a simplistic approximation by assuming 
a single burst population is representative of the entire galaxy, i.e. all stars are
of the same age.  Moreover, the {\it M}/{\it L}$_{\rm I}$ models shown in Figure 4 are 
for stellar masses only, whereas the plotted data are dynamical masses.  Although SSP 
models may be an adequate approximation for a typical quiescent elliptical galaxy, the
models do not take into account the presence of non-stellar matter (e.g. gas, dust, etc), 
nor multiple stellar populations (such as a younger population unseen at {\it I}-band, 
as we posit for LIRGs and ULIRGs).  Both the additional ISM mass and the effects of 
extinction of the {\it I}-band flux from dust will increase the plotted values of 
{\it M}/{\it L}.  Thus, the dynamical {\it M}/{\it L} is almost always 
greater than {\it M}/{\it L} derived from stellar population models, and in the case of 
younger galaxies, the inclusion of two populations brings the models more in line with  
the data \citep[e.g.][]{2001AJ....121.1936G,2006MNRAS.366.1126C,2007iuse.book..107D}.
However, the main point we make here is that at {\it I}-band the ULIRGs appear closer in
mass and age to older gEs (with masses well above {\it m}$^{*}$) in contrast to 
results obtained for the same ULIRGs at near-IR wavelengths.  We do not attempt to derive
absolute stellar population ages for the ULIRGs using a single bandpass, we simply
point out that the comparisons in Figure 4 appear to indicate little or no evidence for
the presence of young stars at {\it I}-band, even though they are clearly present
at other wavelengths.  \\
\indent Separating the two parameters in Figure 4, we now focus exclusively on 
{\it M}$_{\rm Dyn}$ using a standard two-sided KS test.   First, for the QSOs, the Null 
Hypothesis can be rejected (and at the 99$\%$ level) when the {\it M}$_{\rm Dyn}$ 
distributions of the ULIRGs are compared with the RLQs.  Monte Carlo simulations show 
little change, except at the 99$\%$ level.  The Null Hypothesis cannot be rejected 
when the {\it M}$_{\rm Dyn}$ distributions of the ULIRGs are compared with the entire 
sample of QSO hosts.  Next, the Null Hypothesis can be rejected (and at the 99$\%$ level)
when the ULIRGs are compared with the entire sample of SDSS ellipticals.  The only 
qualification is that the Null Hypothesis cannot be rejected for the largest 
$\sigma$$_{\circ}$ bin.   What these results show is that at {\it I}-band the ULIRGs are 
consistent with gEs possibly including RQQ host galaxies. They may not be consistent with 
RLQ host galaxies, but some doubts remain.

\subsubsection{The Dynamical Masses \& Stellar Populations of ULIRGs:  Optical vs. Near-IR}\label{mdyn-i-vs-ir}
\indent As noted earlier, previous dynamical studies of ULIRGs carried out almost
exclusively in the near-IR had concluded that for these systems 
{\it M}$_{\rm Dyn}$ $\leq$ {\it m}$^{*}$.  Figure 5 demonstrates how the 
$\sigma$-Discrepancy in Section \ref{sigmadiscrepancy} can account for the differences
between the results in Sections \ref{fundamental plane} and \ref{mdyn-iband} 
and earlier near-IR only studies.  This figure presents a relative comparison of where 
ULIRGs lie in the {\it M}$_{\rm Dyn}$-{\it M}/{\it L} plane at {\it I}-band ({\it top}) 
and at {\it F160W} ($\sim$ {\it H}-band, {\it bottom}). 
Only the ULIRGs which have kinematic and photometric observations at both wavelengths are 
shown.  The ULIRGs are represented by letters in Figure 5, each letter corresponding to a 
specific galaxy (see the caption for Figure 5 and Table 4).  The two plots share the x-axis 
({\it M}$_{\rm Dyn}$) in order to enable a direct comparison of the mass computed at 
both wavelengths.  The {\it F160W} photometry and kinematic data from the near-IR CO 
band-heads are presented in Table 4.  As in Figure 4, SSP models from M05 are shown with 
Kroupa (solid vector) and Salpeter (light grey) solar metallicity IMFs on the left and 
CB07 for a Chabrier (solid) and Salpeter (light grey) IMFs on the right.  The SSP vectors 
plotted in the {\it F160W} panel were transformed from the original M05 Johnson 
{\it H}-band values to {\it F160W} using $(H-F160W)$ colors computed by processing the 
grid of M05 SEDs at each age with {\tt SYNPHOT}.  At {\it F160W}, the discrepancy between 
the M05 and CB07 models are more pronounced than at {\it I}-band, particularly for the
M05 Kroupa IMF and the CB07 Chabrier and Salpeter IMFs.  The relative ages appear to be 
offset by $\sim$ 0.5 dex between the M05 and CB07.  However, regardless
of the specific age, ULIRGs clearly appear to be older and more massive at {\it I}-band
than at {\it F160W}.  As first noted in Paper I, this effect is not seen in elliptical 
galaxies because they do not have a young stellar population that dominates observations
in any wavelength.

\subsubsection{The Kormendy Relation}\label{kormendy relation}
\indent The analysis presented above is limited by the number of QSO host galaxies
with $\sigma$$_{\circ}$ measurements, particularly RQQs.  The 
$<$$\mu$$>$$_{\rm eff}$-Log {\it R}$_{\rm eff}$ plane (also known as the 
Kormendy Relation) is a photometric projection of the FP for early-type 
systems \cite[e.g.][]{1977ApJ...218..333K,1982modg.proc..113K}, and, like the FP,
is independent of galaxy environment 
\citep[e.g.,][]{1998AJ....116.1591P,2004MNRAS.354..851R,2007A&A...472..773N}.
Although the Kormendy Relation (KR) has significantly more scatter than the FP, 
it is often used as a ``cost-effective'' proxy because the stellar absorption line 
spectroscopy needed to measure $\sigma$$_{\circ}$ can be time-consuming and/or difficult 
to obtain.  This is especially true for QSOs because observations must contend with the 
effects of the bright nucleus which can swamp the underlying host galaxy.  Until recently, 
nearly all studies investigating the dynamical properties of QSO host galaxies and their
relationship to ellipticals, spirals, and mergers have relied solely on photometric 
observations.  Here, the KR is used to increase the number of QSO host galaxies for 
comparison from 6 to 28 RLQs, and from 2 to 25 RQQs.  The main goals of this section are:  
1) to ascertain the positions of advanced ULIRGs with respect to the KR; and 2) to 
determine where the ULIRGs lie relative to the elliptical QSO host galaxies. \\
\indent  Figure 6a ({\it left}) shows the {\it I}-band KR (solid line) derived from the 
entire comparison sample of SDSS ellipticals using the DWLSQ fitting method described 
earlier. The derived {\it I}-band KR is:

\begin{equation} <\mu_{\rm I}>_{\rm eff}\; =\; 17.52^{\pm 0.01}\; +\; 2.26^{\pm 0.01} {\rm Log}\; {\it R}_{\rm eff}.\end{equation}

\noindent with {\it r.m.s.} $=$ 0.39 dex (in units of mag arcsec$^{-1}$) and a 
Pearson Correlation Coefficient of {\it r} $=$ 0.71.  This is similar to the fit from 
\cite{2008A&A...491..731N} for 8,664 early type galaxies at SDSS {\it i}-band 
(z $\leq$ 0.36).  Their slope and intercept is 2.52$^{\pm 0.01}$ and 17.84$^{\pm 0.01}$, 
respectively (transformed to {\it F814W}, the intercept is 17.25), with 
{\it r.m.s.} $=$ 0.42 dex and a Pearson Correlation Coefficient
of {\it r} $=$ 0.72. \\
\indent The ULIRGs, RLQs, RQQs, and SDSS ellipticals are plotted in the {\it left} panel 
of Figure 6a (same symbols as Figure 3). The diagonal dotted shows the locus of an 
{\it L}$^{*}$ galaxy \citep{2003ApJ...592..819B,2003ApJS..149..289B,2010MNRAS.404.1215H}.  
Also shown are the 1, 2, and 3$\sigma$ dispersions of the KR (dotted grey lines).  
As in Figure 3, the SDSS ellipticals follow a gradient in their distribution when grouped 
by $\sigma$$_{\circ}$.  The SDSS ellipticals with the largest $\sigma$$_{\circ}$ lie almost 
exclusively above the KR and the ones with the smallest $\sigma$$_{\circ}$ lie almost 
exclusively below it.  In addition to fitting a KR to the entire SDSS comparison sample, 
fits have been made for each of the SDSS sub-samples (binned by $\sigma$$_{\circ}$) as 
well as the ULIRGs and QSO host galaxies.  The number of objects in each sample, the 
coefficients of the fit, {\it r.m.s.} and Pearson Correlation Coefficient are provided 
for each sample in Table 5.  The slope of the KR does not vary significantly among the 
first three $\sigma$$_{\circ}$ bins, but does decrease significantly for the bin with 
the highest $\sigma$$_{\circ}$. This is somewhat different than the results from 
\cite{2008A&A...491..731N}, which found that the slope of the KR changes significantly 
as a function of luminosity when binned in 1 mag intervals and in intervals of increasing 
luminosity.  \\
\indent The {\it r.m.s.} of the 8 ULIRGs is 1.01, within 2.5$\sigma$ of the KR.  However, 
IRAS 19542+1110 is a significant outlier from the {\it I}-band KR ($\sim$ 6$\sigma$).  
Excluding it, the {\it r.m.s.} of the ULIRGs decreases to 0.52 ($\sim$ 1.3$\sigma$ from
the KR).  Once again, a 2D KS test was used to compare the distribution of the ULIRGs
with all of the SDSS ellipticals.  Here, the results do change when IRAS 19542+1110 is 
included or excluded.  When included, the Null Hypothesis cannot be rejected.  When
it is excluded, the Null Hypothesis can be rejected at the 95$\%$ level.  However, 
when the ULIRGs are compared with the SDSS ellipticals in the 165-225 km s$^{-1}$ and
225-420 km s$^{-1}$ bins, then the Null Hypothesis cannot be rejected, whether or not 
IRAS 19542+1110 is excluded.   This implies that the ULIRGs are always consistent with 
the most massive SDSS ellipticals, and the inclusion of IRAS 19542+1110 also makes 
them consistent with a broader range of ellipticals.  \\
\indent When the ULIRGs are compared with the RLQs, RQQs, and the QSO host 
galaxies taken together as one sample, the 2D KS test shows that in all cases the Null 
Hypothesis cannot be rejected.  Although the majority of ULIRGs and QSOs fall within 
3$\sigma$ of the KR, they lie systematically above the relation.  
The computed slopes and intercepts are listed in Table 5.
Figure 6b ({\it right panel}) shows the best-fit KRs for the ULIRGs, RLQs, 
and RQQs from Table 5.  The fits are plotted as shaded regions which account for the
1$\sigma$ errors in both slope and intercept.\\
\indent The ULIRG and QSO fits have steeper slopes than the SDSS ellipticals 
(including the sub-samples binned by $\sigma$$_{\circ}$).  The fits to the ULIRGs and 
QSOs are consistent with each other. Figure 6b shows the considerable overlap in 
KR parameter space among ULIRGs, RLQs, and RQQs. Similarly, \cite{2009ApJ...701..587V} 
found at {\it F160W} that the slope of the KR fits to {\it PSF}-subtracted PG QSOs and 
ULIRGs (in which the bulk of the star-formation should have been removed in the 
{\it PSF}-subtraction), along with QSOs from D03 and \cite{2008ApJ...678...22H},
transformed assuming $(R-H)$ $=$ 2.8, were indistinguishable from each other,
but still significantly steeper than inactive ellipticals.  The QSO results in Table 5 
and Figure 6b 
also match results from earlier attempts to place QSO host galaxies on the KR at optical
wavelengths \citep[e.g.][D03,F04, see bottom of Table 5]{2002ApJ...580...96O}.\\
\indent A standard two-sided KS comparison of each individual parameter in the KR
($<$$\mu$$>$$_{\rm eff}$ and Log {\it R}$_{\rm eff}$) between the ULIRGs and the RLQs, 
RQQs, and the two QSO samples together produces the same result in all cases: the 
Null Hypothesis cannot be rejected.  Given that the ULIRGs are now compared with much 
larger samples of RLQs and RQQs, Figure 6 presents stronger support than the FP or 
{\it M}$_{\rm Dyn}$-{\it M}/{\it L} plane for the assertion that ULIRGs are consistent 
with the host galaxies of QSOs, both as a single population, and when divided
into RLQs and RQQs.  \\
\indent Finally, since the RLQ and RQQ photometric samples are sufficiently large, 
one can compare them with each other using both the 2D KS test and the 1D comparison for 
each parameter in the KR.  Here, the results are quite interesting because the Null 
Hypothesis cannot be rejected for the 2D case nor can it be rejected for 
$<$$\mu$$>$$_{\rm eff}$.  However, for Log {\it R}$_{\rm eff}$ the comparison is less 
clear.  Using the modified KS test, the Null Hypothesis can be rejected at the 95$\%$ 
level, but using the standard KS formula for large samples it cannot be 
rejected.  Such a difference between the methods suggests one should err on the side
of caution in making any strong statements about whether the RLQs and RQQs are 
significantly different in a statistical sense.  Both D03 and F04 reached a similar 
conclusion, noting that the mean {\it R}$_{\rm eff}$ of RLQs and RQQs are the same 
within the 1$\sigma$ errors.  These results raise doubt about the intrinsic 
differences between RLQs and RQQs.  This reinforces the need for more kinematic
($\sigma$$_{\circ}$) measurements for RLQs and RQQs.  

\section{Discussion}\label{discussion}
\indent The goal of this paper has been to use the first results of the {\it I}-band
dynamical survey of advanced ULIRGs to address  two key questions:  
1) Does the $\sigma$-Discrepancy extend to the more luminous 
ULIRG population?  and 2)  At {\it I}-band are the dynamical properties of advanced ULIRGs 
consistent with gEs, including the host galaxies of QSOs?  Here, we briefly discuss
the implications for the results presented so far.

\subsection{The $\sigma$-Discrepancy}
\indent The $\sigma$-Discrepancy does appear to extend to ULIRGs such that
$\sigma$$_{\circ, \rm CaT}$ $>$ $\sigma$$_{\circ, \rm CO}$.
Moreover, the results here agree with what was posited in
Paper I, namely the correlation between Log {\it L}$_{\rm IR}$ and the 
$\sigma$-Discrepancy, and in turn, the predicted range of $\sigma$$_{\circ, \rm CaT}$ 
values for ULIRGs.  Just as with the LIRGs in Paper I, the ULIRGs can be described 
as Janus-like.  Like the Roman deity, they present two different faces depending on how 
they are viewed. At {\it I}-band the face of an old stellar population is observed, 
while at near-infrared wavelengths the face of a young stellar population dominates 
(see Figure 19 in Paper I). Since in Paper I, and in this work, we are observing 
{\it only} single nuclei mergers, the luminosity evolution of the disks in these systems 
may be nearly monotonic and decreasing.  In this subclass of (U)LIRGs, the lower the 
{\it L}$_{\rm IR}$, the further along will be important processes such as feedback that 
clears out the star-forming ISM 
\citep[e.g.][]{2010A&A...518L..41F,2010A&A...518L.155F,2011ApJ...733L..16S}
and the subsequent aging of the starburst.  As the YCSD becomes fainter, its dominance 
in both photometric and kinematic measurements in the near-IR subsides, reducing the 
observed $\sigma$-Discrepancy.  As the lack of this discrepancy in bonafide ellipticals 
demonstrates \citep[Paper I,][]{2011MNRAS.412.2017V} at some point in the evolutionary 
sequence $\sigma$$_{\circ, optical}$ $=$ $\sigma$$_{\circ, near-IR}$.  Part of this 
explanation is an oversimplification because it assumes that every merger reaches a 
ULIRG stage and that LIRGs are stages before and after the luminosity peak.  On the other 
hand, regardless of whether all LIRG mergers will be or at some point have been ULIRGs, 
lower {\it L}$_{\rm IR}$ means a less luminous YCSD, and a less dusty nuclear region.  

\subsection{The Evolution of ULIRGs into QSOs}
\indent The extension of the $\sigma$-Discrepancy to ULIRGs leads directly to the second 
question; whether the structure and kinematics of ULIRGs are consistent with those of gEs, 
including QSO host galaxies.  We address this in two steps, beginning with a comparison 
to elliptical galaxies in general. Previous results from near-IR stellar kinematics and 
photometry
\citep[e.g.][]{2001ApJ...563..527G,2002ApJ...580...73T,2004ApJ...613..781D,2005ApJ...621..725C,2006ApJ...651..835D}
concluded ULIRGs were the progenitors of low-intermediate mass ellipticals 
($<$ {\it m}$^{*}$) based on two arguments.  The first was simply that the
measured values of $\sigma$$_{\circ}$ obtained from near-IR stellar lines were
significantly smaller than those of typical or giant ellipticals obtained from optical
stellar absorption lines (e.g. Ca H\&K, \ion{Mg}{Ib}, CaT).  This was based on the 
assumption that the near-IR stellar absorption lines or stellar band-heads 
probe the global properties of the ULIRGs.  The results presented here and in Paper I 
imply this is not the case.  The second is comparing the values of {\it M}$_{\rm Dyn}$, 
computed from the CO $\sigma$$_{\circ}$ and near-IR photometry, with some fiducial stellar 
mass representative of the stellar mass function of galaxies.  The earlier studies above   
all used {\it m}$^{*}$ $=$ 7.07$\times$10$^{10}$ {\it h}$^{-2}$ {\it M}$_{\odot}$
(or 1.25$\times$10$^{11}$ {\it M}$_{\odot}$ for the cosmology used here) from
\cite{2001MNRAS.326..255C}.
In other words, because in the near-IR {\it M}$_{\rm Dyn}$ $<$ {\it m}$^{*}$, ULIRGs 
cannot form gEs, let alone an average elliptical.  They must be the 
progenitors of low-intermediate mass ellipticals.  However, the value of {\it m}$^{*}$
used for comparison is actually the larger of two possible values from 
\cite{2001MNRAS.326..255C}.  The other is 
{\it m}$^{*}$ $=$ 3.43$\times$10$^{10}$ {\it h}$^{-2}$ {\it M}$_{\odot}$
(6.0$\times$10$^{10}$ {\it M}$_{\odot}$ for the cosmology used here).
The larger value comes from using a Salpeter IMF which over-predicts the amount of 
low-mass stars, rather than a Kennicutt or ``diet'' Salpeter IMF which compensates for 
this effect \citep[see Section 6.1 of][]{2003ApJS..149..289B,2001ApJ...550..212B}.
A variety of methods have converged towards 
{\it m}$^{*}$ $\sim$ 3$\times$10$^{10}$ {\it M}$_{\odot}$ 
(the value used here) which also appears to be the transition region between the blue 
cloud and red sequence and the threshold above which AGN activity is more likely to be found
\citep[e.g.][]{2001ApJ...550..212B,2003ApJS..149..289B,2003MNRAS.341...54K,2004ApJ...600..681B,2007ApJ...663..834B,2008MNRAS.388..945B}.
In other words, the claim that (U)LIRGs form sub-{\it m}$^{*}$ ellipticals is partly
based on the selection of the larger of two possible {\it m}$^{*}$ values. In the bottom
panel of Figure 5 the ULIRGs straddle the {\it m}$^{*}$ line.  This remains the same when
near-IR data for other advanced ULIRGs from  
\cite{2001ApJ...563..527G,2002ApJ...580...73T,2006ApJ...651..835D} are considered.
Instead, the real question raised by these earlier results is why the observed 
$\sigma$$_{\circ, \rm CO}$ of ULIRGs and near-IR half-light radii were inconsistent with 
observations at other wavelengths, including molecular gas masses, star-formation rates, 
high-velocity outflows, etc.  Similar to the results for non-IR luminous mergers and 
LIRGs (e.g. RJ06a, Paper I), the kinematic and photometric properties of the ULIRGs 
measured at {\it I}-band are now statistically consistent with {\it m}$^{*}$ and 
larger ellipticals, including gEs.  \\
\indent QSOs and ULIRGs have a great number of similarities, including:  
bolometric luminosities and space densities out at least z $\sim$ 0.4 
\citep{1986ApJ...303L..41S,2001ApJ...555..719C};
an overlap in the FIR-radio correlation 
\citep[e.g.][]{1996ARAA..34..749S,2001ApJ...554..803Y}; H$_{\rm 2}$ masses 
of $\sim$ 10$^{9-10}$ {\it M}$_{\odot}$ 
\citep[e.g.][]{1989A&A...213L...5S,1997A&A...318...15C,2001AJ....121.1893E,2003ApJ...585L.105S,2007A&A...470..571B,2009ASPC..408...35E,2009AJ....138..262E,2012ApJ...750...92X};
post-starburst stellar populations in or near the nucleus along with
tidal tails and peculiar morphologies indicative of a relatively recent gas-rich merging event 
\citep[e.g.][]{1984ApJ...283...64M,1986ApJ...311..526H,1988AJ.....96.1575H,1992AJ....104....1H,2006ApJS..166...89G,2007ApJ...669..801C,2008ApJ...677..846B,2011MNRAS.410.1550R,2011MNRAS.412..960T};
and an overlap in the distribution  of {\it L}$_{\rm IR}$ 
(12.24 $\pm$ 0.44, 12.25 $\pm 0.47$, and 12.17 $\pm$ 0.16 for RLQs, RQQs, and ULIRGs,
respectively), in which the Null Hypothesis cannot be rejected.
However, the sizes and masses of ULIRGs and QSO host galaxies have previously been
reported as significantly different.  
The {\it I}-band dynamical results presented here alleviate this discrepancy.
There is now a much stronger dynamical link between ULIRGs and QSO host galaxies.  
The strongest result comes from the KR, 
in part, due to the large sample size of QSOs. The 2D KS test for the distribution of 
objects in the Log {\it R}$_{\rm eff}$-$<$$\mu$$_{\rm I}$$>$$_{\rm eff}$ plane, as well 
as the two-sided KS tests for each parameter rule out statistical differences between 
ULIRGs and QSOs (whether grouped together or compared separately as RLQs and RQQs).  
While past comparisons have relied on the KR to reject the notion that ULIRGs evolve
into QSO host galaxies, the same comparison here at {\it I}-band strongly supports the
S88 paradigm.\\
\indent However, it is still important to compare the kinematics of ULIRGs and QSOs.  
Unfortunately, the tradeoff for doing so is the significantly smaller sample size of QSOs, 
including the loss of comparing ULIRGs with only RQQs.  The results for the FP are 
consistent with those of the KR.  QSOs 
(either taken together or just RLQs) and ULIRGs show no statistical difference in their 
distribution in FP parameter space.  It is only when the parameter 
$\sigma$$_{\circ}$ is considered alone that things become less clear.  There is a weak 
statistical difference between RLQs and ULIRGs for this parameter (see Table C1), 
primarily due to the large RLQ errors.  In the case of {\it M}$_{\rm Dyn}$ the differences
are stronger and the RLQ $\sigma$$_{\circ}$ errors do not affect the results significantly 
(see Table C1).  How can these results be reconciled with those from the KR (and its
individual parameters) which uses the larger photometric QSO sample?  
Is $\sigma$$_{\circ}$ really different for RLQs and ULIRGs, or does it 
appear to be different because the 6 RLQs in the kinematic sub-sample happen to be
non-representative of the larger RLQ sample?  To test this, the 2D and standard 
two-sided KS tests were re-run for the KR and its parameters between the ULIRGs and the 
{\it kinematic} QSO sub-sample only.  Just one difference emerges from the results 
listed in Table C1.  The Null Hypothesis {\it can} be rejected for the 
Log {\it R}$_{\rm eff}$ comparison between ULIRGs and the 6 RLQs.  This explains 
the rejection of the Null Hypothesis for {\it M}$_{\rm Dyn}$ 
between RLQs and ULIRGs because $\sigma$$_{\circ}$ and Log {\it R}$_{\rm eff}$ are used
to compute the mass.  Although the kinematic results for the RLQs are uncertain, when 
the QSO hosts are considered as a single population, it suggests kinematic similarities 
exist between them and ULIRGs.  Overall, these results clearly demonstrate the need for 
more measurements of $\sigma$$_{\circ}$ in RLQs and RQQs to confirm the results from the 
KR and probe whether the kinematic differences between RLQs and ULIRGs are real 
(and if any exist between ULIRGs and RQQs). \\
\indent Although beyond the scope of this paper, these results also raise a conundrum
in regards to whether RQQs and RLQs are dynamically different.  The Null Hypothesis can 
be rejected when their luminosities are compared.  This implies that their host masses are
different (assuming some {\it M}/{\it L} transformation from stellar population models 
and that the ages of the stellar populations in RLQs and RQQs are the same).  However, 
their 2D distributions in the KR are not statistically different 
(nor are the two parameters when each is compared separately).  Since the KR is a 
projection of the FP, which in turn is related to the correlation between {\it M} and 
{\it M}/{\it L}, it implies that RLQs and RQQs are {\it not} dynamically different. 

\section{Summary \& Future Work}
\noindent The main results of this paper are summarized below.\\
\indent {\bf 1)} The $\sigma$-Discrepancy, first reported in RJ06a and quantified in 
Paper I for LIRGs is shown to extend to ULIRG luminosities.  
The $\sigma$$_{\circ}$ measured from the CaT stellar absorption lines are systematically 
larger than those obtained from the near-IR CO band-heads.  With the addition of ULIRGs 
the correlation between $\sigma$$_{\rm Frac}$ and Log {\it L}$_{\rm IR}$
remains unchanged from Paper I.  We posit that for the single nuclei (U)LIRGs presented
here and in Paper I, this relationship results from feedback processes that cause 
monotonic aging and dimming of the YCSD population and the clearing out of the dusty, 
star-forming medium.\\
\indent {\bf 2)} At {\it I}-band, ULIRGs are nearly an order of magnitude more massive 
than previously measured in the near-IR, and are consistent with ellipticals ranging
from {\it m}$^{*}$ to gEs. All of the ULIRGs presented here lie closer to 
the Fundamental Plane and Kormendy Relation than in near-IR studies.  \\
\indent {\bf 3)} At {\it I}-band, the {\it M}/{\it L} values of ULIRGs appear to indicate 
the presence of an old, evolved stellar population, similar to quiescent ellipticals.  
Yet in the near-IR, ULIRGs reflect much younger populations, matching the well established 
observations of significant quantities of molecular gas and high rates of SFR. \\
\indent {\bf 4)} At {\it I}-band ULIRGs are dynamically similar to QSO host galaxies, 
further supporting the S88 paradigm.  ULIRGs are statistically consistent with 
the positions of both RLQs and RQQs on the Fundamental Plane and Kormendy Relation.  
This result uses the same methods of comparison which in the past have been used to 
demonstrate that ULIRGs do not evolve into QSO host galaxies. However, when the 
ULIRGs are compared with the kinematic sub-sample of RLQs, there is a statistical
difference (the Null Hypothesis can be rejected) for $\sigma$$_{\circ}$, 
{\it M}$_{\rm Dyn}$, and Log {\it R}$_{\rm eff}$.  The impact of this difference is 
weakened by two caveats; the large errors in $\sigma$$_{\circ}$ for the RLQs which
affects the KS tests (see Table C1;) and the contradiction which arises from the
Null Hypothesis not being rejected for ULIRGs and RLQs when the full sample of 28 RLQs
are considered.  These results demonstrate the need for more $\sigma$$_{\circ}$ 
measurements for RLQs (to resolve the contradiction) and RQQs in order to make a viable 
kinematic comparison with ULIRGs and confirm the results presented here.\\
\indent {\bf 5)} Finally, an homogenized {\it I}-band sample of RLQ and RQQ host galaxies 
are presented here which can be used for future dynamical studies.  The QSO hosts
can be used as either a control sample for further comparisons with IR-luminous systems
or as a representative sample itself for future studies of QSO host galaxies.\\
\indent  These are the first results from a much broader survey to establish an accurate 
mass distribution for ULIRGs and to re-evaluate how these systems fit into the broader 
picture of the formation and evolution of elliptical galaxies and QSOs.  It is clear that 
more kinematic (stellar $\sigma$$_{\circ}$) observations are needed for RLQ and RQQ host 
galaxies in order to confirm the results presented here.  However, the results presented 
so far are consistent with both the Toomre Hypothesis and QSO evolution scheme presented
in S88.  Future work will focus on a multi-wavelength approach which will continue to 
use optical observations to measure {\it M}$_{\rm Dyn}$ and near-IR observations to probe 
the central kpc in ULIRGs.  A comparison of central black hole masses 
({\it M}$_{\bullet}$) will also be made, in order to determine the location of ULIRGs with
respect to the {\it M}$_{\bullet}$-$\sigma$ relation.  These will require observations to 
infer {\it M}$_{\bullet}$ independent of $\sigma$ or host luminosity.  Assessing detailed
and accurate properties of local ULIRGs is key to formulating a better understanding of 
similar systems (e.g. dust obscured galaxies, sub-millimeter galaxies, etc..) at higher 
redshifts. 

\acknowledgments
Some of this research was performed while B. Rothberg held a National Research
Council Associateship Award at the Naval Research Laboratory.  Basic research in astronomy 
at the Naval Research Laboratory is funded by the Office of Naval Research.   This work 
was supported in part by a NASA Keck PI Data Award, administered by the NASA 
Exoplanet Science Institute. Data presented herein were obtained at the W. M. Keck 
Observatory from telescope time allocated to the National Aeronautics and Space 
Administration through the agency’s scientific partnership with the California Institute 
of Technology and the University of California. The Observatory was made possible 
by the generous financial support of the W. M. Keck Foundation.
The authors would like to thank: Drs. Sylvain Veilleux and David Rupke for providing 
information and help with the flux calibration; Drs. David Floyd and Timothy Hamilton 
for providing additional information and data; and Drs. Gabriela Canalizo and Scott Dahm
for helpful discussions.  The authors would also like to thank the anonymous referee
for useful and detailed comments which strengthened the paper.  
This research made use of the OSX Version of SCISOFT assembled by Dr. Nor Pirzkal.
This work has made use of the EZGal Model Generator (http://www.baryons.org/ezgal/).
We would also like to acknowledge the University of Maximegalon's MISPWOSO.  
This research has made use of the NASA/IPAC Extragalactic Database (NED) which is 
operated by the Jet Propulsion Laboratory, California Institute of Technology, under 
contract with the National Aeronautics and Space Administration.  This publication makes 
use of data products from the Wide-field Infrared Survey Explorer, which is a joint 
project of the University of California, Los Angeles, and the Jet Propulsion 
Laboratory/California Institute of Technology, funded by the National Aeronautics and 
Space Administration. This work made use of the Sloan Digital Sky Survey (SDSS).  
Funding for the SDSS and SDSS-II has been provided by the Alfred P. Sloan Foundation, 
the Participating Institutions, the National Science Foundation, the U.S. Department of 
Energy, the National Aeronautics and Space Administration, the Japanese Monbukagakusho, 
the Max Planck Society, and the Higher Education Funding Council for England. 
The SDSS Web Site is http://www.sdss.org/.  The SDSS is managed by the Astrophysical 
Research Consortium for the Participating Institutions.   The Participating Institutions 
are the American Museum of Natural History, Astrophysical Institute Potsdam, 
University of Basel, University of Cambridge, Case Western Reserve University, 
University of Chicago, Drexel University, Fermilab, the Institute for Advanced Study, 
the Japan Participation Group, Johns Hopkins University, the Joint Institute for Nuclear 
Astrophysics, the Kavli Institute for Particle Astrophysics and Cosmology, 
the Korean Scientist Group, the Chinese Academy of Sciences (LAMOST), Los Alamos 
National Laboratory, the Max-Planck-Institute for Astronomy (MPIA), 
the Max-Planck-Institute for Astrophysics (MPA), New Mexico State University, 
Ohio State University, University of Pittsburgh, University of Portsmouth, 
Princeton University, the United States Naval Observatory, and the University of 
Washington.

\clearpage
\begin{deluxetable}{lccccc}
\tabletypesize{\footnotesize}
\tablewidth{0pt}
\tablenum{1}
\pagestyle{empty}
\tablecaption{ULIRG Galaxy Sample}
\tablecolumns{6}
\tablehead{
\colhead{Galaxy} &
\colhead{R.A.} &
\colhead{Dec.} &
\colhead{{\it z}} &
\colhead{Log {\it L}$_{\rm IR}$}&
\colhead{{\it E}$(B-V)$}\\
\colhead{Name} &
\colhead{(J2000)}&
\colhead{(J2000)}&
\colhead{} &
\colhead{({\it L}$_{\odot}$)}&
\colhead{(mag)}
}
\startdata
IRAS F02021-2103   &02 04 27 &-20 49 41 &0.116 &12.02\tablenotemark{a,b}        &0.020\\
IRAS 05189-2524    &05 21 01 &-25 21 45 &0.043 &12.11\tablenotemark{a}          &0.026\\
IRAS F10378+1108   &10 40 29 &+10 53 18 &0.136 &12.26\tablenotemark{a,b}        &0.029\\
IRAS 11387+4116    &11 41 22 &+40 59 51 &0.148 &12.03\tablenotemark{b,c,d}      &0.015\\
IRAS 12540+5708    &12 56 14 &+56 52 25 &0.042 &12.48\tablenotemark{a}          &0.009\\
IRAS 17208-0014    &17 23 21 &-00 17 01 &0.043 &12.34\tablenotemark{a}          &0.304\\
IRAS 19542+1110    &19 54 35 &+11 19 02 &0.064 &12.03\tablenotemark{a,b}        &0.199\\
IRAS 23365+3604    &23 39 01 &+36 21 09 &0.064 &12.13\tablenotemark{a,b}        &0.096\\
\enddata
\tablecomments{{\footnotesize R.A. is in units of hours, minutes, and seconds, and Dec is in units of degrees, arcminutes, and arcseconds. 
{\it L}$_{\rm IR}$ is the  8-1000 $\micron$  total flux measured from the 12, 25, 60, and 100 $\micron$ IRAS passbands. 
(a) All or some IRAS fluxes from \cite{2003AJ....126.1607S};
(b) 12$\micron$ flux from {\it WISE};
(c) All or some IRAS fluxes from \cite{1990IRASF.C......0M};
(d) 22$\micron$ flux from {\it WISE}.}}
\end{deluxetable}

\begin{deluxetable}{llcccc}
\tabletypesize{\footnotesize}
\tablewidth{0pt}
\tablenum{2}
\pagestyle{empty}
\tablecaption{ULIRG Observation Log}
\tablecolumns{6}
\tablehead{
\colhead{Galaxy} &
\colhead{Imaging} &
\colhead{Imaging} &
\colhead{ESI} &
\colhead{ESI} &
\colhead{ESI} \\
\colhead{Name} &
\colhead{Camera/} &
\colhead{Integration Time} &
\colhead{Integration Time} &
\colhead{P.A.} &
\colhead{Slitwidth} \\
\colhead{} &
\colhead{Filter} &
\colhead{(sec)} &
\colhead{(sec)} &
\colhead{(deg)} &
\colhead{(arcsec)} 
}
\startdata
IRAS F02021-2103\tablenotemark{a}         &{\it HST} WFPC2/F814W\tablenotemark{c}    &800   &3600        &55.0    &1.0   \\
IRAS 05189-2524\tablenotemark{b}          &{\it HST} ACS/F814W\tablenotemark{d}      &730   &900         &0.0     &1.0    \\
IRAS F10378+1108\tablenotemark{b}         &SDSS/Sloan z                              &54.1  &1800        &0.0     &1.0    \\
IRAS 11387+4116\tablenotemark{b}          &SDSS/Sloan z                              &54.1  &1800        &0.0     &1.0    \\
IRAS 12540+5708\tablenotemark{a}          &{\it HST} ACS/F814W\tablenotemark{d}      &830   &900         &45.0    &0.75   \\
IRAS 17208-0014\tablenotemark{a}          &{\it HST} ACS/F814W\tablenotemark{d},NIC2/F160W\tablenotemark{e}      &720,224   &3900        &40.0    &1.0    \\
IRAS 19542+1110\tablenotemark{a}          &{\it HST} ACS/F814W\tablenotemark{d}      &720   &3600        &0.0     &1.0    \\
IRAS 23365+3604\tablenotemark{a}          &{\it HST} ACS/F814W\tablenotemark{d},NIC2/F160W\tablenotemark{f}      &750,2496   &5400        &300.0   &1.0    \\
\enddata
\tablecomments{{\footnotesize(a) ESI P.I. Rothberg; (b) ESI P.I. Sanders;
(c) {\it HST} Program 6346, P.I. Borne; (d) {\it HST} Program 10592, P.I. Evans;
(e) {\it HST} Program 7219, P.I. Scoville; (f) {\it HST} Program 11235, Surace.
The WFPC2 observations were centered on the WF3 chip.}}
\end{deluxetable}

\begin{deluxetable}{llllllll}
\tabletypesize{\footnotesize}
\tablewidth{0pt}
\tablenum{3}
\pagestyle{empty}
\tablecaption{ULIRG Rest-frame {\it I}-band Properties}
\tablecolumns{8}
\tablehead{
\colhead{Galaxy} &
\colhead{{\it M}$_{\rm I}$} &
\colhead{{\it R}$_{\rm eff}$} &
\colhead{$<$$\mu$$_{\rm I}$$>$$_{\rm eff}$}&
\colhead{CaT {$\sigma$$_{\circ}$}} &
\colhead{CaT {\it V$_{\odot}$}} &
\colhead{CaT} &
\colhead{{\it M}$_{\rm dyn}$}\\
\colhead{Name} &
\colhead{(mag)} &
\colhead{(kpc)} &
\colhead{(mag arcsec$^{-2}$)}&
\colhead{(km s$^{-1}$)} &
\colhead{(km s$^{-1}$)} &
\colhead{Template Star} &
\colhead{($\times$10$^{11}${\it M}$_{\odot}$)}
}
\startdata
IRAS F02021-2103   &-23.61$^{\pm 0.05}$ &9.51$^{\pm 0.23}$ &19.84$^{\pm 0.05}$ &209\tablenotemark{a}$^{\pm 8}$  &34679$^{\pm 8}$ &G8III HD 35369                 &5.79$^{\pm 0.46}$\\
IRAS 05189-2524    &-22.71$^{\pm 0.09}$ &2.59$^{\pm 0.10}$ &17.93$^{\pm 0.12}$ &265\tablenotemark{a}$^{\pm 7}$  &12869$^{\pm 7}$ &K0III HD 206067                &2.54$^{\pm 0.15}$\\
IRAS F10378+1108   &-22.96$^{\pm 0.10}$ &3.09$^{\pm 0.14}$ &18.05$^{\pm 0.23}$ &280\tablenotemark{b}$^{\pm 11}$ &41007$^{\pm 9}$ &G1III $\alpha$ Sge (HD 185758) &3.38$^{\pm 0.31}$\\
IRAS 11387+4116    &-23.01$^{\pm 0.01}$ &3.22$^{\pm 0.07}$ &18.09$^{\pm 0.14}$ &198\tablenotemark{b}$^{\pm 9}$  &44575$^{\pm 7}$ &M0III $\lambda$ Dra (HD 100029) &1.76$^{\pm 0.16}$\\
IRAS 12540+5708    &-23.42$^{\pm 0.10}$ &5.88$^{\pm 0.27}$ &18.98$^{\pm 0.13}$ &346\tablenotemark{c}$^{\pm 9}$  &12584$^{\pm 10}$ &G5II HD 36079                 &9.86$^{\pm 0.68}$\\
IRAS 17208-0014    &-22.88$^{\pm 0.11}$ &9.37$^{\pm 0.50}$ &20.54$^{\pm 0.12}$ &261\tablenotemark{a}$^{\pm 5}$  &12798$^{\pm 5}$ &K0III HD 206067                &8.90$^{\pm 0.58}$\\
IRAS 19542+1110    &-23.52$^{\pm 0.09}$ &0.77$^{\pm 0.03}$ &14.43$^{\pm 0.14}$ &169\tablenotemark{a}$^{\pm 6}$  &18718$^{\pm 5}$ &G8III HD 35369                 &0.31$^{\pm 0.02}$\\
IRAS 23365+3604    &-23.16$^{\pm 0.08}$ &4.16$^{\pm 0.19}$ &18.50$^{\pm 0.10}$ &221\tablenotemark{a}$^{\pm 6}$  &19310$^{\pm 5}$ &K1.5III $\alpha$ Boo (HD 124897) &2.83$^{\pm 0.19}$\\
\enddata
\tablecomments{{\footnotesize
Fluxes are in VEGA magnitudes.  The values have also been corrected 
for Galactic Reddening using dust maps from \cite{1998ApJ...500..525S} 
and scaling from Table 6 in \cite{2011ApJ...737..103S}, 
assuming {\it R}$_{\rm V}$ $=$ 3.1 \citep{1999PASP..111...63F}.   
The {\it A}$_{\lambda}$ scaling factors used were:  
{\it A}$_{\rm F814W}$ (ACS/WFC) $=$ 1.52;
{\it A}$_{\rm F814W}$ (WFPC2) $=$ 1.54; and 
{\it A}$_{\rm z}$ $=$ 1.26.
Circular isophotes were used for the {\it F814W} photometry.  Elliptical
isophotes were transformed to equivalent radii for SDSS {\it z}-band 
photometry on IRAS F10378+1108 and IRAS 11387+4116.
(a) Measured in 1.53 kpc diameter aperture; 
(b) Corrected to 1.53 kpc diameter;
(c) Measured 2.08 kpc NW from the nucleus and corrected to a 1.53 kpc diameter.}}
\end{deluxetable}

\clearpage
\begin{deluxetable}{lllll}
\tabletypesize{\footnotesize}
\tablewidth{0pt}
\tablenum{4}
\pagestyle{empty}
\tablecaption{Near-IR ULIRG Properties}
\tablecolumns{5}
\tablehead{
\colhead{Galaxy} &
\colhead{CO $\sigma$$_{\circ}$}&
\colhead{{\it M}$_{\rm Dyn}$ via CO}& 
\colhead{{\it R}$_{\rm eff}$ ({\it F160W})} &
\colhead{{\it M}$_{\rm 160W}$ } \\
\colhead{Name} &
\colhead{(km s$^{-1}$)} &
\colhead{($\times$10$^{11}${\it M}$_{\odot}$)}&
\colhead{(kpc)} &
\colhead{(mag)} 
}
\startdata
IRAS F02021-2103 ({\tt A}) &143$^{\pm 21}$\tablenotemark{a}  &1.09             &3.85\tablenotemark{d}             &-24.89\tablenotemark{f}\\
IRAS 05189-2524  ({\tt B}) &131$^{\pm 16}$\tablenotemark{a}  &0.11             &0.48\tablenotemark{d}             &-23.96\tablenotemark{f}\\
IRAS 12540+5708  ({\tt C}) &117$^{\pm 10}$\tablenotemark{b}  &0.20             &1.05\tablenotemark{d}              &-24.22\tablenotemark{f}\\
IRAS 17208-0014  ({\tt D}) &223$^{\pm 15}$\tablenotemark{c}  &0.94$^{\pm 0.14}$  &1.36$^{\pm 0.10}$\tablenotemark{e} &-24.43$^{\pm 0.18}$\\
IRAS 23365+3604  ({\tt E}) &143 $^{\pm 15}$\tablenotemark{c} &0.57$^{\pm 0.13}$  &2.00$^{\pm 0.22}$\tablenotemark{e} &-25.25$^{\pm 0.41}$\\
\enddata
\tablecomments{{\footnotesize The letters {\tt A}-{\tt E} correspond to objects in Figure 5.
Fluxes are in VEGA magnitudes.  The values have also been corrected 
for Galactic Reddening using dust maps from \cite{1998ApJ...500..525S} 
and scaling of {\it A}$_{\rm F160W}$ $=$ 0.51, 
assuming {\it R}$_{\rm V}$ $=$ 3.1 \citep{1999PASP..111...63F}.   
All values of $\sigma$$_{\circ, CO}$ have been corrected to a common aperture of 1.53 kpc.
Some values in the table do not include errors because they were not available from the source materials.
(a) \cite{2006ApJ...651..835D}, M0III template star used (HD 99817);
(b) \cite{2002ApJ...580...73T}  K5Ib template star used (HD 200576);
(c) \cite{2001ApJ...563..527G}  M0III template star used (HD 99817);
(d) PSF-subtracted S\'{e}rsic fit ({\it n} $=$ 4) \citep{2006ApJ...643..707V} 
from elliptical isophotes converted to equivalent radii using ellipticities
derived from their Table 2;
(e) Photomery measured using circular isophotes.
(f) \cite{2006ApJ...643..707V}.
The values of $\sigma$$_{\circ, CO}$ in Column 2 and {\it M}$_{\rm Dyn}$ values computed in 
Column 3 are only for comparison purposes with the CaT derived properties in Table 3
and do not represent the global properties of the systems.}}
\end{deluxetable}

\clearpage
\begin{deluxetable}{llllll}
\tabletypesize{\footnotesize}
\tablewidth{0pt}
\tablenum{5}
\pagestyle{empty}
\tablecaption{Kormendy Relation Fits}
\tablecolumns{5}
\tablehead{
\colhead{Sample} &
\colhead{{\it N}} &
\colhead{Slope} &
\colhead{Intercept} &
\colhead{{\it r.m.s.}}&
\colhead{{\it r}}
}
\startdata   
\cutinhead{Rest-frame {\it I}-band}
SDSS Ellipticals                                                 &9255   &2.26$^{\pm 0.01}$  &17.56$^{\pm 0.01}$  &0.35  &0.71\\
SDSS Ellipticals 85 $\leq$ $\sigma_{\circ}$ $\leq$ 125 km s$^{-1}$ &1656   &2.76$^{\pm 0.01}$  &17.75$^{\pm 0.01}$  &0.38  &0.76\\
SDSS Ellipticals 125 $<$ $\sigma$$_{\circ}$ $\leq$ 165 km s$^{-1}$ &3301   &2.69$^{\pm 0.01}$  &17.33$^{\pm 0.01}$  &0.36  &0.82\\
SDSS Ellipticals 165 $<$ $\sigma$$_{\circ}$ $\leq$ 225 km s$^{-1}$ &3524   &2.72$^{\pm 0.01}$  &17.02$^{\pm 0.01}$  &0.35  &0.85\\
SDSS Ellipticals 225 $<$ $\sigma$$_{\circ}$ $\leq$ 420 km s$^{-1}$ &774    &2.37$^{\pm 0.02}$  &17.00$^{\pm 0.01}$  &0.21  &0.87\\
ULIRGs\tablenotemark{a}                                          &8      &4.50$^{\pm 0.14}$  &15.68$^{\pm 0.09}$  &0.22  &0.98\\
RLQs\tablenotemark{b}                                            &28     &3.51$^{\pm 0.52}$  &15.71$^{\pm 0.45}$  &0.54  &0.79\\
RQQs\tablenotemark{b}                                            &25     &3.40$^{\pm 0.44}$  &16.11$^{\pm 0.33}$  &0.56  &0.84\\
All QSO Host Galaxies\tablenotemark{b}                           &53     &3.25$^{\pm 0.32}$  &16.07$^{\pm 0.26}$  &0.57  &0.81\\
\cutinhead{Restframe {\it I}-band for Previously Published Samples\tablenotemark{c}}
RLQs (O'Dowd et al. 2002)                                        &16     &3.81$^{\pm 0.66}$\tablenotemark{d}  &15.37$^{\pm 0.59}$  &0.46  &0.83\\
RLQs (D03)                                                       &10     &3.66$^{\pm 0.47}$\tablenotemark{e}  &15.77$^{\pm 0.42}$  &0.23  &0.93\\
RLQs (D03, excluding two outliers\tablenotemark{f})              &8      &3.05$^{\pm 0.30}$\tablenotemark{g}  &16.42$^{\pm 0.29}$  &0.12  &0.97\\
RQQs (D03)                                                       &9      &3.40$^{\pm 0.73}$\tablenotemark{h}  &16.43$^{\pm 0.58}$  &0.37  &0.86\\
RLQs (F04)                                                       &7      &3.68$^{\pm 0.10}$\tablenotemark{i}  &15.83$^{\pm 0.08}$  &0.37  &0.94\\
RQQs (F04)                                                       &7      &3.13$^{\pm 0.20}$\tablenotemark{i}  &15.86$^{\pm 0.15}$  &0.45  &0.91\\
RLQs + RQQs Combined (F04)                                       &14     &3.20$^{\pm 0.07}$\tablenotemark{j}  &16.00$^{\pm 0.06}$  &0.48  &0.91\\
\enddata
\tablecomments{{\footnotesize As a comparison, the earlier {\it B}-band and {\it V}-band
KRs had slopes of 3.02  \citep{1977ApJ...218..333K} and 2.94 \citep{1987IAUS..127..379H},
repsectively, using the surface brightness {\it at} the effective radius.  Two fits 
each were made to the RLQ, RQQ, and RLQ+RQQ data to account for the absence of errors
in the D03 sample.  If errors were available a DWLSQ fit, which takes into account errors 
in X and Y, was used.  Otherwise, the data points were equally weighted and a standard 
least-squares fit was used. The fits plotted in Figure 6 include all data for the ULIRGs, RLQs, and RQQs.
(a) IRAS 19542+1110 is considered an outlier because it is 6.3$\sigma$ from the 
KR plotted in Figure 6, while all other ULIRGs are $\leq$ 2$\sigma$ from the line.  
Excluding it changes the fit to:
$<$$\mu$$>$$_{\rm eff}$ 3.85$^{\pm 0.19}$$\times$Log {\it R}$_{\rm eff}$ + 16.19$^{\pm 0.14}$, {\it r.m.s.}$=$ 0.24 and {\it r} $=$0.96;
(b) No errors are available for the D03 sample.  The KR was fit using a simplified least-squares fit
with equal weighting of errors for the dependent data points.  When the D03 data are excluded and a DWLSQ method was used
the fits change to:
RLQs ({\it N} $=$ 22):  $<$$\mu$$>$$_{\rm eff}$ 3.58$^{\pm 0.07}$$\times$Log {\it R}$_{\rm eff}$ + 15.66$^{\pm 0.06}$, {\it r.m.s.}$=$ 0.57 and {\it r} $=$0.81;
RQQs ({\it N} $=$ 20):  $<$$\mu$$>$$_{\rm eff}$ 3.79$^{\pm 0.09}$$\times$Log {\it R}$_{\rm eff}$ + 15.15$^{\pm 0.06}$, {\it r.m.s.}$=$ 0.83 and {\it r} $=$0.83;
All QSOs ({\it N} $=$ 42):  $<$$\mu$$>$$_{\rm eff}$ 4.23$^{\pm 0.05}$$\times$Log {\it R}$_{\rm eff}$ + 14.85$^{\pm 0.03}$, {\it r.m.s.}$=$ 0.74 and {\it r} $=$0.82;
(c) {\it All} of the QSOs used to derive the preivously published KRs are 
also part of the QSO Comparison Sample in Appendix B.  The previously published fits used 
the surface brightness {\it at} the effective radius, not the surface 
brightness {\it within} the effective radius ($<$$\mu$$>$$_{\rm eff}$), as used here.  The original fits used
different cosmologies, {\it k}-corrections, transformations to different rest-frame filters,
and different fitting techniques.  They have been refit here to provide a clean comparison;
(d) Original published restframe {\it R}-band published slope $=$ 2.75$^{\pm 1.2}$;
(e) Original published observed {\it F675W}-band published slope $=$ 3.98$^{\pm 0.71}$;
(f) PG 1004+130 and OX 169;
(g) Original published restframe {\it R}-band published slope $=$ 3.19$^{\pm 0.67}$;
(h) {\it F675W}-band published slope $=$ 2.99$^{\pm 0.34}$;
(i) RLQ and RQQ samples from F04 were not fit separately;
(j) Original published restframe {\it V}-band published slope $=$ 3.33$^{\pm 0.7}$.}}
\end{deluxetable}

\clearpage
{\begin{figure}
\plotone{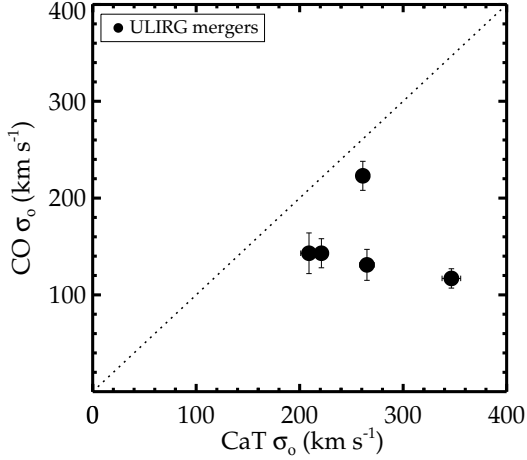}
\caption{Comparison of the central velocity dispersions ($\sigma$$_{\circ}$) 
measured from the Calcium II Triplet (CaT) stellar absorption lines at 
$\lambda$ $\sim$ 0.85$\micron$ (x-axis) and those measured from the CO band-head at 
1.6 or 2.3$\micron$ (y-axis) for 5 ULIRGs.} 
\end{figure}}

{\begin{figure}
\plotone{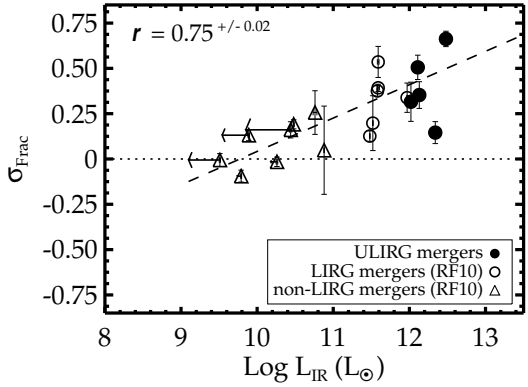}
\caption{Values of $\sigma$$_{\rm Frac}$, which quantifies the relative difference between
$\sigma$$_{\circ, \rm CaT}$ and $\sigma$$_{\circ, \rm CO}$ for each advanced merger, are compared 
with the independently measured quantity Log {\it L}$_{\rm IR}$.  The horizontal dotted 
line represents $\sigma$$_{\rm Frac}$ $=$ 0 and the diagonal dashed line shows the 
weighted least-squares fit to the data.   Also shown is the Pearson Correlation 
Coefficient ({\it r}).  Values of  {\it r} range from -1 
(strong anti-correlation) to +1 (strong correlation).  There is a strong correlation 
between $\sigma$$_{\rm Frac}$ and Log {\it L}$_{\rm IR}$.} 
\end{figure}}

{\begin{figure}
\plotone{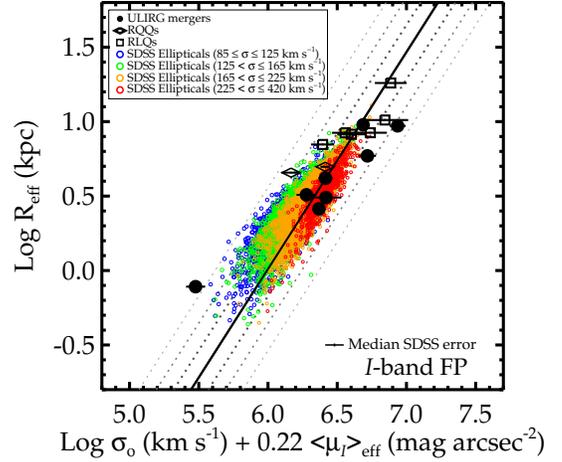}
\caption{The diagonal solid line plotted in this figure represents the {\it I}-band 
Fundamental Plane \citep{2009MNRAS.396.1171H} seen edge-on.  It is transformed from the 
Sloan {\it i}-band filter to our preferred cosmology.  Plotted in this figure are:  
advanced ULIRGs (filled circles), the host galaxies 
of Radio Loud (open squares), and Radio Quiet (open flattened diamonds) QSOs, and the 
SDSS comparison sample of 9,255 ellipticals (open circles in four colors).  In this, 
and subsequent plots, the SDSS ellipticals are shown binned into four groups based on 
$\sigma$$_{\circ}$:
1) 85 km s$^{-1}$ $\leq$ $\sigma$$_{\circ}$ $\leq$ 125 km s$^{-1}$ (1656 blue);
2) 120 km s$^{-1}$ $<$ $\sigma$$_{\circ}$ $\leq$ 165 km s$^{-1}$ (3301 green);
3) 165 km s$^{-1}$ $<$ $\sigma$$_{\circ}$ $\leq$ 225 km s$^{-1}$ (3524 orange);
and 4) 225 km s$^{-1}$ $<$ $\sigma$$_{\circ}$ $\leq$ 420 km s$^{-1}$ (774 red).
The ULIRGs and QSO host galaxies have values of $\sigma$$_{\circ}$ which would 
place them in either the 3rd or 4th bin.
Error bars are plotted for the ULIRG sample and QSOs (where available), along with the
median errors for the SDSS sample.  Over-plotted are the 1, 2, and 3$\sigma$ scatter 
of the FP (dotted diagonal lines in dark to light grey).} 
\end{figure}}

{\begin{figure}
\plotone{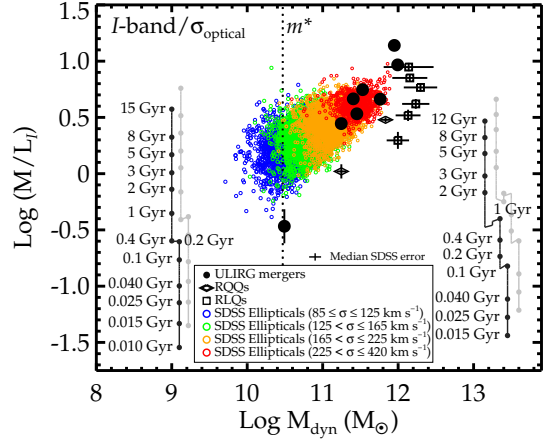}
\caption{This figure compares ({\it M}$_{\rm Dyn}$) and {\it M}/{\it L}
values at {\it I}-band. Plotted here are the ULIRGs, SDSS comparison ellipticals and 
QSO host galaxies (same symbols used in Figure 3). The over-plotted solid vectors in the 
panel show the temporal evolution of {\it M}/{\it L} for a burst single stellar population
(SSP) with solar metallicity and a Kroupa (solid black line) or Salpeter (grey line) 
IMF (M05). The value of {\it M} for the SSPs represents the stellar mass ({\it m}$^{*}$)
while the masses plotted for the galaxies are the total virial dynamical mass
({\it M}$_{\rm Dyn}$).  The horizontal placement of the {\it M}/{\it L} 
vectors are for display only.   The vertical dotted line represents the mass of an 
{\it m}$^{*}$ galaxy ($\sim$ 3$\times$10$^{10}$ {\it M}$_{\odot}$). Error bars are 
plotted for the ULIRG sample and QSOs (where available), along with the median errors 
for the SDSS sample.}
\end{figure}}
\clearpage

{\begin{figure}
\plotone{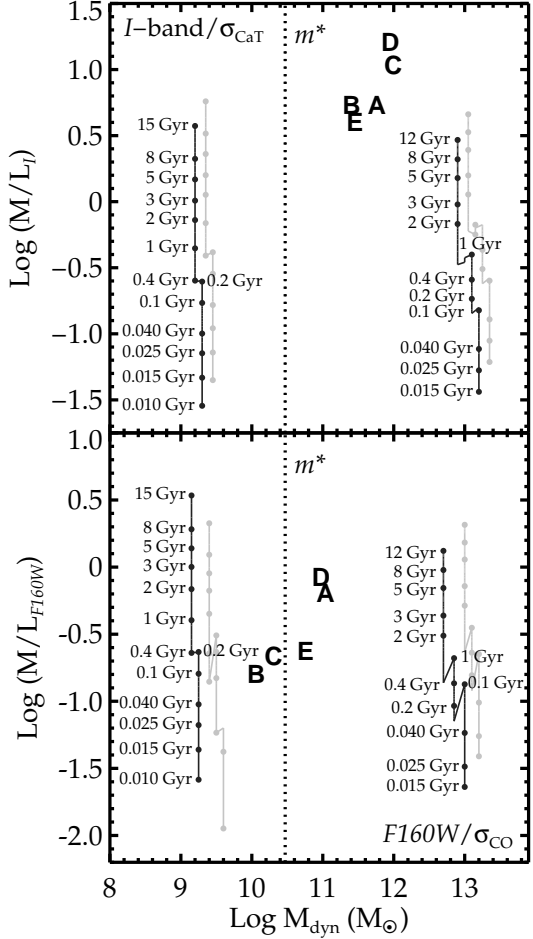}
\caption{Shown here is a plot similar to Figure 4, comparing the same set of ULIRGs at 
{\it I}-band ({\it top}) with their parameters at {\it F160W} ({\it bottom}).  
The {\it top} panel uses {\it I}-band photometry along with $\sigma$$_{\circ, \rm CaT}$ and 
the {\it bottom} panel uses HST/{\it NIC2} {\it F160W} photometry with 
$\sigma$$_{\circ, \rm CO}$. The 5 ULIRGs plotted in both panels are:  
{\sf A} $=$ IRAS F02021-2103, {\sf B} $=$ IRAS 05189-2524, {\sf C} $=$ IRAS 12540+5708, 
{\sf D} $=$ IRAS 17208-0014, and {\sf E} $=$ IRAS 23365+3604.  The solid vectors in  
both panels represent the temporal evolution of {\it M}/{\it L} at {\it I}-band 
({\it top}) and {\it F160W} ({\it bottom}) for a SSP with 
solar metallicity and a Kroupa (solid black line) or Salpeter (grey line) IMF 
(M05). Once again, the {\it M} from the SSPs represents the stellar mass ({\it m}$^{*}$),
while for the plotted galaxies it represents {\it M}$_{\rm Dyn}$.
The horizontal placement of the {\it M}/{\it L} vectors in both panels are for 
display only.  The vertical dotted line in both panels represents an 
{\it m}$^{*}$ galaxy ($\sim$ 3$\times$10$^{10}$ {\it M}$_{\odot}$).  In the bottom panel
ULIRGs {\tt A}, {\tt B}, and {\tt C} have been {\it PSF}-subtracted 
(FWHM $=$ 0.14 arcsec pixel$^{-1}$ corresponding to 0.27, 0.11, and 0.10 kpc for
{\tt A}, {\tt B}, and {\tt C}, respectively). This has likely removed some of the 
star-formation contribution to the galaxy luminosity.  Therefore, their vertical positions 
in the bottom panel are upper limits suggesting their ages and {\it M}/{\it L} may be 
smaller than what is shown. }
\end{figure}}

{\begin{figure}
\epsscale{1.25}
\plotone{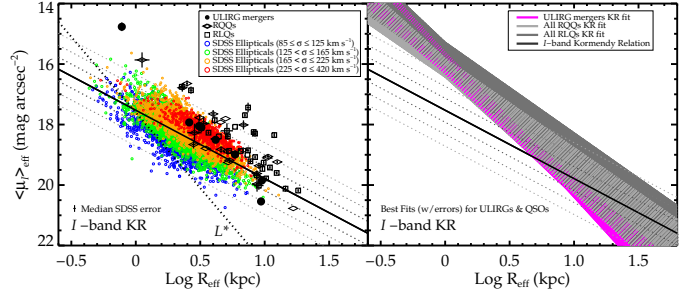}
\caption{Shown in both panels is the Kormendy Relation (KR), a photometric projection of the 
Fundamental Plane. The {\it left} panel compares 
photometric properties of the sample of 8 advanced ULIRGs with a larger sample of 28 RLQ 
and 25 RQQ host galaxies.  The diagonal solid black line plotted in this figure represents 
the {\it I}-band KR computed from the SDSS comparison sample of 9,255 Ellipticals
({\it r.m.s.} $=$ 0.39 dex).  Error bars are plotted for the ULIRG sample and QSOs 
(where available), along with the median errors for the SDSS sample.  The diagonal 
dashed line shows the range of $<$$\mu$$_{\rm I}$$>$ and Log {\it R}$_{\rm eff}$ for 
an {\it L}$^{*}$ galaxy of constant luminosity.  The dotted grey lines (from dark to 
light) represent 1, 2, and 3$\times${\it r.m.s.} of the KR. The {\it r.m.s.} of the
various samples are: ULIRGs $=$ 1.01 (0.52 without IRAS 19542+1110), RLQs $=$ 0.95,
RQQs $=$ 0.91, all QSOs $=$ 0.93.
Plotted in the {\it right} panel are the derived fits to the ULIRGs, RLQs, and RQQs 
from Table 5.  The filled regions for each sample display the fit including
the $\pm$1$\sigma$ errors in the slope and intercept.  The diagonal solid black and dotted
grey lines represent the KR and the 1-3$\times${\it r.m.s.} from the fit. }
\end{figure}}

\appendix
\setcounter{figure}{0} 
\setcounter{table}{0}
\section{Sloan Digital Sky Survey Comparison Sample Selection Criteria}
\indent The selection criteria used to generate the SDSS comparison sample are described 
below.  They are based upon the criteria used in 
\cite{2003AJ....125.1817B,2009MNRAS.394.1978H,2009MNRAS.396.1171H} 
to select early-type galaxies (including S0 or lenticulars) for deriving the Fundamental 
Plane in several bandpasses. The criteria are divided into photometric (items 1-7) and 
spectroscopic (items 8-12) categories. The selection criteria required that the elliptical 
galaxy must be present in both the photometric ({\tt PhotoObjAll}) and spectroscopic 
({\tt SpecObj}) databases.   Only Sloan {\it i}-band photometry was used 
for the photometric catalog.  No magnitude or flux limitations were imposed on the 
sample.\\  
\indent The biggest differences between the selection criteria used here and those in 
\cite{2003AJ....125.1817B,2009MNRAS.394.1978H,2009MNRAS.396.1171H}
are: redshift (z $\leq$ 0.15 here vs. 0.36); restricting the sample to elliptical 
morphologies only (affected by {\tt lnlDev} and {\tt lnlExp}; and
$\sigma$ $\geq$ 85 km s$^{-1}$ vs 60 km s$^{-1}$ to avoid the effects of instrumental 
resolution (see Appendix B in \cite{2003AJ....125.1817B}).  No parameters or values
for those parameters were selected that would exclude any elliptical in the comparison 
sample from being a part of the HB09 sample.\\
\indent 1) {\tt MODE} $=$ 1 selects the primary photometry for each object, rejecting possible duplications as 
wells as objects flagged as saturated, near the edge of a CCD. \\
\indent 2) {\tt PARENTID} $=$ 0 and {\tt nCHILD} $=$ 0 rejects galaxies blended with other objects, or the child of a 
de-blended set of objects.  Given the spatial resolution of the photometry and spectral resolution of the data, 
de-blended superimposed objects may yield unrealistic measured parameters. Not setting these parameters resulted 
in a number of double counted galaxies in which the same $\sigma$ is assigned to both parent and child galaxy, 
producing spurious results.\\
\indent 3) {\tt FRACDEV} $=$ 1:  defined as the fraction of total flux which contributes to a de Vaucouleurs 
\citep{1953MNRAS.113..134D} or {\it r}$^{1/4}$ fit to the galaxy light profile.\\
\indent 4) {\tt devAB} $\geq$ 0.6:  defined as the axis ratio of the minor to major axis for a de Vaucouleurs fit. 
Analysis by \cite{2009MNRAS.394.1978H,2009MNRAS.396.1171H} demonstrated that while {\tt FRACDEV} $=$ 1 eliminates most disk galaxies 
from SDSS photometric catalogs, a non-trivial amount remain.  Values of {\tt devAB} $\geq$ 0.6 significantly 
improve the removal of late-type galaxies.\\
\indent 5) {\tt Type} $=$ 3:  Morphological classification as a galaxy.\\
\indent 6) {\tt lnlExp} \& {\tt lnlDev} $>$ -9999:  {\tt lnlExp} and {\tt lnlDev} are maximum likelihood functions 
which estimates the best-fit model parameters convolved with an estimate of the seeing.  Smaller values indicate 
a larger likelihood.  The criteria cutoff assure that the measured values have meaning, as values of -9999 mean 
no data is available.\\
\indent 7) {\tt lnlDev} at least 10$\%$  $>$ {\tt lnlExp}:  This criteria selects objects in which the likelihood 
of a de Vaucouleurs fit is 10$\%$ greater than an exponential fit.  This is a recommended setting from SDSS for 
selecting elliptical galaxies.\\
\indent 8) {\tt specObjID} $\ne$ 0: reject objects without spectroscopic observations.\\
\indent 9) {\tt SpecClass} $=$ 2:  Spectral classification as a galaxy.\\
\indent 10) {\tt eClass} $<$ 0: a one-dimensional classification of spectral type from Principal Component Analysis 
\citep{2004AJ....128..585Y} in which negative values correspond to absorption line galaxies with old stellar 
populations and positive values correspond to star-forming galaxies.\\
\indent 11) {\tt sn0}, {\tt sn1}, and {\tt sn2} $\geq$ 10.0:    
The values of $\sigma$ in DR7 are measured by fitting the rest-frame wavelength range 0.4-0.7 $\micron$.  
The {\tt sn0}, {\tt sn1}, and {\tt sn2} criteria were selected to ensure $\sigma$ was measured from spectra with 
sufficient S/N.\\
\indent 12) $\sigma$ $\geq$ 85 km s$^{-1}$ and $<$ 420 km s$^{-1}$:  The SDSS spectra are re-sampled to a dispersion 
of log $\lambda$ = 10$^{-4}$ dex pixel$^{-1}$ which corresponds to 69 km s$^{-1}$.  The actual spectral resolution 
varies from 85-105 km s$^{-1}$ for galaxies in the SDSS spectra \citep{2003AJ....125.1817B} due to variations
as a function of wavelength.  The DR7 (and DR6) $\sigma$ values differ from earlier data releases in that 
they are measured using a direct-fitting method \citep{1992MNRAS.254..389R} with the assumption of a Gaussian 
profile, rather than a Fourier fitting routine. The latter appears to bias $\sigma$'s $<$
 150 km s$^{-1}$ systematically higher 
\citep{2007AJ....133.1954B}.  The direct-fitting method is the same method used for measuring the 
$\sigma$$_{\circ}$ from the spectra of the ULIRGs presented here although the profile shape is fit with a 
Gauss-Hermite polynomial rather than a Gaussian because the S/N is higher  (see Section 3, RJ06a, and Paper I).  
The DR7 release notes also warn that $\sigma$ $>$ 420 km s$^{-1}$ are not reliable.  To err
on the side of caution we have selected the lower cutoff to be the approximate limit of,
rather than below the instrumental resolution.\\
\indent Finally, for completeness, we apply a rest-frame radius correction for each elliptical galaxy 
in the SDSS comparison sample.  Since early-type galaxies have color gradients, yielding slightly larger
{\it R}$_{\rm eff}$ at bluer wavelengths, we use equation 6 from \cite{2009MNRAS.394.1978H},
which interpolates the observed radii from adjacent bands.  The median correction is
+0.012 kpc.

\section{Data for the Comparison Samples of RLQ \& RQQ Host Galaxies}
\clearpage
\begin{deluxetable}{lccclccccc}
\tabletypesize{\footnotesize}
\tablewidth{0pt}
\tablenum{1}
\pagestyle{empty}
\tablecaption{QSO Comparison Sample}
\tablecolumns{10}
\tablehead{
\colhead{Galaxy} &
\colhead{R.A.} &
\colhead{Dec.} &
\colhead{{\it z}} &
\colhead{Log {\it L}$_{\rm IR}$}&
\colhead{{\it E}$(B-V)$\tablenotemark{a}} &
\colhead{Camera/} &
\colhead{{\it M}$_{\rm I}$} &
\colhead{{\it R}$_{\rm eff}$} &
\colhead{$<$$\mu$$_{\rm I}$$>$$_{\rm eff}$}\\
\colhead{Name} &
\colhead{(J2000)}&
\colhead{(J2000)}&
\colhead{} &
\colhead{({\it L}$_{\odot}$)} & 
\colhead{(mag)}&
\colhead{Filter} &
\colhead{(mag)} &
\colhead{(kpc)} &
\colhead{(mag arcsec$^{-2}$)}
}
\startdata
\cutinhead{Radio Loud QSO Comparison Sample}
HB89 0031-707    &00 34 05  &-70 25 52  &0.363   &12.22\tablenotemark{b,c,d,3,4}   &0.031  &WF3/{\it F791W}\tablenotemark{$\dagger$}  &-23.79$^{\pm 0.08}$       &7.74$^{\pm 0.28}$       &19.21$^{\pm 0.08}$\\
HB89 0110+297    &01 13 24  &+29 58 15  &0.363   &12.47\tablenotemark{b,c,d,3,4}   &0.063  &WF2/{\it F814W}\tablenotemark{$\dagger$}  &-23.57$^{\pm 0.14}$       &8.66$^{\pm 0.56}$       &19.67$^{\pm 0.12}$\\
3C48             &01 37 41  &+33 09 35  &0.367   &13.03\tablenotemark{b,1,2,3,4}   &0.044  &PC/{\it F814W}\tablenotemark{$\ast$}      &-25.59$^{\pm 0.03}$       &11.85$^{\pm 0.10}$      &18.34$^{\pm 0.10}$\\
PKS 0137+012     &01 39 57  &+01 31 46  &0.260   &12.54\tablenotemark{b,c,3,4}     &0.029  &WF2/{\it F675W}\tablenotemark{$\ast$}     &-24.26$^{\pm 0.03}$       &9.81$^{\pm 0.13}$       &19.26$^{\pm 0.07}$\\
PKS 0202-76      &02 02 13  &-76 20 03  &0.389   &12.66\tablenotemark{b,d,3,4}     &0.051  &PC/{\it F702W}\tablenotemark{$\ast$}      &-23.84$^{\pm 0.07}$       &2.68$^{\pm 0.08}$       &16.84$^{\pm 0.12}$\\
3C59             &02 07 09  &+29 31 41  &0.109   &(11.75)\tablenotemark{b}         &0.063  &WF2/{\it F675W}\tablenotemark{$\ddagger$} &-23.31                  &5.28                  &18.86\\
PKS 0312-77      &03 11 55  &-76 51 51  &0.223   &11.92\tablenotemark{b,c,4}       &0.097  &PC/{\it F702W}\tablenotemark{$\ast$}      &-24.35$^{\pm 0.02}$       &10.98$^{\pm 0.07}$      &19.42$^{\pm 0.08}$\\
PKS 0736+01      &07 39 18  &+01 37 05  &0.191   &11.80\tablenotemark{e,4}         &0.128  &WF2/{\it F675W}\tablenotemark{$\ast$}     &-24.12$^{\pm 0.02}$       &8.42$^{\pm 0.09}$       &19.07$^{\pm 0.06}$\\
PKS 0812+02      &08 15 22  &+01 55 00  &0.402   &12.44\tablenotemark{b,c,3,4}     &0.029  &WF2/{\it F814W}\tablenotemark{$\dagger$}  &-24.43$^{\pm 0.03}$       &12.39$^{\pm 0.21}$      &19.60$^{\pm 0.05}$\\
PKS 0903+16      &09 06 31  &+16 46 12  &0.412   &12.83\tablenotemark{b,c,d}       &0.040  &PC/{\it F814W}\tablenotemark{$\ast$}      &-23.89$^{\pm 0.24}$       &5.09$^{\pm 0.55}$       &18.20$^{\pm 0.25}$\\
PG 1004+130      &10 07 26  &+12 38 56  &0.240   &12.26\tablenotemark{c,g,4}       &0.040  &WF2/{\it F675W}\tablenotemark{$\ddagger$} &-24.34                  &5.88                  &18.07\\
PKS 1020-103     &10 22 32  &-10 37 44  &0.190   &11.17\tablenotemark{c,d,e}       &0.046  &WF2/{\it F675W}\tablenotemark{$\ddagger$} &-23.39                  &4.40                  &18.39 \\
PKS 1058+110     &11 00 47  &+10 46 13  &0.422   &11.41\tablenotemark{c,d,e}       &0.026  &WF2/{\it F814W}\tablenotemark{$\dagger$}  &-23.32$^{\pm 0.18}$       &9.01$^{\pm 0.75}$       &20.01$^{\pm 0.14}$\\
HB89 1150+497    &11 53 24  &+49 31 09  &0.333   &12.29\tablenotemark{c,d,f,4}     &0.021  &WF2/{\it F814W}\tablenotemark{$\dagger$}  &-23.74$^{\pm 0.08}$       &5.18$^{\pm 0.18}$       &18.39$^{\pm 0.11}$\\
HB89 1208+322    &12 10 37  &+31 57 06  &0.389   &12.31\tablenotemark{b,c,d,3}     &0.017  &WF3/{\it F791W}\tablenotemark{$\dagger$}  &-23.11$^{\pm 0.03}$       &3.66$^{\pm 0.05}$       &18.26$^{\pm 0.05}$\\
PKS 1217+02      &12 20 11  &+02 03 42  &0.240   &12.22\tablenotemark{b,d,4}       &0.022  &WF2/{\it F675W}\tablenotemark{$\ddagger$} &-23.89                  &7.35                  &19.00\\
PG 1226+023      &12 29 06  &+02 03 09  &0.158   &12.72\tablenotemark{g}           &0.021  &WF3/{\it F606W}\tablenotemark{$\ast$}     &-24.89$^{\pm 0.04}$       &8.41$^{\pm 0.13}$       &18.29$^{\pm 0.10}$\\
PKS 1233-24      &12 35 37  &-25 12 17  &0.355   &12.46\tablenotemark{b,c,d}       &0.097  &WF2/{\it F814W}\tablenotemark{$\dagger$}  &-23.60$^{\pm 0.07}$       &2.27$^{\pm 0.07}$       &16.75$^{\pm 0.05}$\\
PG 1302-102      &13 05 33  &-10 33 19  &0.278   &12.35\tablenotemark{c,d,h}       &0.043  &WF3/{\it F606W}\tablenotemark{$\ast$}     &-24.89$^{\pm 0.04}$       &8.21$^{\pm 0.15}$       &18.24$^{\pm 0.10}$\\
PG 1425+267      &14 27 35  &+26 32 14  &0.366   &12.64\tablenotemark{d,i,3,4}     &0.019  &WF3/{\it F814W}\tablenotemark{$\ast$}     &-24.25$^{\pm 0.03}$       &14.50$^{\pm 0.20}$      &20.12$^{\pm 0.10}$\\
PG 1545+210      &15 47 43  &+20 52 17  &0.264   &11.95\tablenotemark{c,g,4}       &0.042  &WF3/{\it F606W}\tablenotemark{$\ast$}     &-24.06$^{\pm 0.05}$       &7.42$^{\pm 0.17}$       &18.85$^{\pm 0.09}$\\
PG 1704+608      &17 04 41  &+60 44 31  &0.372   &12.71\tablenotemark{g,4}         &0.069  &PC/{\it F702W}\tablenotemark{$\ast$}      &-25.48$^{\pm 0.07}$       &7.43$^{\pm 0.20}$       &17.43$^{\pm 0.16}$\\
PKS 2135-14      &21 37 45  &-14 32 56  &0.200   &12.06\tablenotemark{b,c,d,4}     &0.050  &WF2/{\it F675W}\tablenotemark{$\ast$}     &-23.84$^{\pm 0.03}$       &10.25$^{\pm 0.16}$      &19.78$^{\pm 0.07}$\\
OX 169           &21 43 35  &+17 43 49  &0.211   &11.99\tablenotemark{b,c,d,3,4}   &0.111  &WF2/{\it F675W}\tablenotemark{$\ast$}     &-23.77$^{\pm 0.03}$       &4.09$^{\pm 0.02}$       &17.85$^{\pm 0.13}$\\
4C +31.63        &22 03 15  &+31 45 38  &0.295   &12.82\tablenotemark{b,c,d,3,4}   &0.124  &PC/{\it F702W}\tablenotemark{$\ast$}      &-25.11$^{\pm 0.03}$       &7.02$^{\pm 0.09}$       &17.68$^{\pm 0.08}$\\
PKS 2247+14      &22 50 25  &+14 19 52  &0.237   &12.18\tablenotemark{b,c,3,4}     &0.050  &WF2/{\it F675W}\tablenotemark{$\ast$}     &-24.01$^{\pm 0.02}$       &9.59$^{\pm 0.08}$       &19.46$^{\pm 0.07}$ \\
PG 2349-014      &23 51 56  &-01 09 13  &0.174   &11,81\tablenotemark{c,d,g}       &0.027  &WF2/{\it F675W}\tablenotemark{$\ast$}     &-24.68$^{\pm 0.02}$       &18.18$^{\pm 0.12}$      &20.18$^{\pm 0.06}$\\
PKS 2355-082     &23 58 09  &-08 00 04  &0.210   &11.86\tablenotemark{b,c,d,3,4}   &0.040  &WF2/{\it F675W}\tablenotemark{$\ddagger$} &-23.76                  &6.50                  &18.87\\
\cutinhead{Radio Quiet QSO Comparison Sample}
LBQS 0020+0018   &00 23 11  &+00 35 18  &0.423   &12.33\tablenotemark{b,c,d,3}     &0.024  &PC/{\it F675W}\tablenotemark{$\ast$}      &-23.59$^{\pm 0.12}$       &2.29$^{\pm 0.12}$       &16.78$^{\pm 0.13}$\\
HB89 0054+144    &00 57 09  &+14 46 10  &0.171   &12.36\tablenotemark{b,d,4}       &0.046  &WF3/{\it F606W}\tablenotemark{$\ast$}     &-23.92$^{\pm 0.04}$       &4.54$^{\pm 0.05}$       &17.92$^{\pm 0.13}$\\
LBQS 0100+0205   &01 03 13  &+02 21 10  &0.393   &12.62\tablenotemark{b,c,d,3,4}   &0.021  &PC/{\it F675W}\tablenotemark{$\ast$}      &-23.23$^{\pm 0.23}$       &3.35$^{\pm 0.34}$       &17.96$^{\pm 0.24}$\\
Mrk 1014         &01 59 50  &+00 23 41  &0.164   &12.59\tablenotemark{g}           &0.029  &WF2/{\it F675W}\tablenotemark{$\ddagger$} &-24.56                  &10.45                 &19.10\\
HB89 0244+194    &02 47 40  &+19 40 58  &0.176   &11.59\tablenotemark{b,c,d,3,4}   &0.110  &WF2/{\it F675W}\tablenotemark{$\ast$}     &-23.01$^{\pm 0.06}$       &5.31$^{\pm 0.14}$       &19.18$^{\pm 0.09}$\\ 
HS 0624+6907     &06 30 02  &+69 05 04  &0.370   &12.62\tablenotemark{b,c,d,3,4}   &0.098  &WF3/{\it F791W}\tablenotemark{$\dagger$}  &-24.73$^{\pm 0.16}$       &6.84$^{\pm 0.49}$       &18.01$^{\pm 0.15}$\\
MS 0754.6+3928   &07 58 00  &+39 20 39  &0.096   &11.69\tablenotemark{c,d,j,4}     &0.066  &PC/{\it F814W}\tablenotemark{$\ast$}      &-23.91$^{\pm 0.04}$       &2.49$^{\pm 0.04}$       &16.64$^{\pm 0.10}$\\
PG 0923+201      &09 25 54  &+19 54 05  &0.190   &12.07\tablenotemark{c,d,j}       &0.042  &WF3/{\it F606W}\tablenotemark{$\ast$}     &-23.64$^{\pm 0.05}$       &8.69$^{\pm 0.20}$       &19.62$^{\pm 0.10}$\\
PG 0953+415      &09 56 52  &+41 15 22  &0.234   &11.82\tablenotemark{c,d}         &0.012  &WF2/{\it F675W}\tablenotemark{$\ddagger$} &-22.89                  &5.20                  &19.25\\
PG 1001+291      &10 04 02  &+28 55 35  &0.329   &12.80\tablenotemark{b,3,4}       &0.022  &WF3/{\it F791W}\tablenotemark{$\dagger$}  &-23.98$^{\pm 0.08}$       &9.01$^{\pm 0.35}$       &19.36$^{\pm 0.10}$\\
PG 1012+008      &10 14 54  &+00 33 37  &0.186   &11.93\tablenotemark{c,d,3,4}     &0.031  &WF2/{\it F675W}\tablenotemark{$\ddagger$} &-23.90                  &16.66                 &20.77\\
He 1029-1401     &10 31 54  &-14 16 51  &0.086   &11.35\tablenotemark{b,c,4}       &0.067  &WF3/{\it F606W}\tablenotemark{$\ast$}     &-23.24$^{\pm 0.02}$       &5.19$^{\pm 0.04}$       &18.90$^{\pm 0.08}$\\
PG 1202+28       &12 04 42  &+27 54 12  &0.165   &11.72\tablenotemark{c,g}         &0.021  &WF3/{\it F606W}\tablenotemark{$\ast$}     &-23.27$^{\pm 0.03}$       &3.12$^{\pm 0.04}$       &17.77$^{\pm 0.10}$\\
PG 1216+069      &12 19 20  &+06 38 39  &0.331   &12.53\tablenotemark{c,d,i,3,4}   &0.022  &PC/{\it F702W}\tablenotemark{$\ast$}      &-23.30$^{\pm 0.18}$       &8.72$^{\pm 0.70}$       &19.96$^{\pm 0.25}$\\
LBQS 1230+0947   &12 33 25  &+09 31 23  &0.414   &13.02\tablenotemark{b,c,3,4}     &0.021  &WF3/{\it F791W}\tablenotemark{$\dagger$}  &-23.92$^{\pm 0.08}$       &3.98$^{\pm 0.13}$       &17.65$^{\pm 0.13}$\\
EQS B1252+0200   &12 55 19  &+01 44 12  &0.345   &12.59\tablenotemark{b,d,3,4}     &0.018  &WF3/{\it F791W}\tablenotemark{$\dagger$}  &-22.57$^{\pm 0.28}$       &2.85$^{\pm 0.36}$       &18.27$^{\pm 0.30}$\\
EQS B1254+0206   &12 57 06  &+01 50 39  &0.421   &12.71\tablenotemark{b,c,3,4}     &0.020  &WF2/{\it F814W}\tablenotemark{$\dagger$}  &-24.64$^{\pm 0.04}$       &10.62$^{\pm 0.22}$      &19.05$^{\pm 0.04}$\\
EQS B1255-0143   &12 58 15  &-01 59 19  &0.410   &12.39\tablenotemark{b,c,d,3,4}   &0.018  &WF2/{\it F814W}\tablenotemark{$\dagger$}  &-22.94$^{\pm 0.29}$       &1.11$^{\pm 0.14}$       &15.86$^{\pm 0.25}$\\
PG 1307+085      &13 09 47  &+08 19 48  &0.155   &11.73\tablenotemark{c,d,g,4}     &0.033  &WF3/{\it F606W}\tablenotemark{$\ast$}     &-23.05$^{\pm 0.05}$       &4.58$^{\pm 0.10}$       &18.82$^{\pm 0.11}$\\
PG 1416-129      &14 19 03  &-13 10 44  &0.129   &11.49\tablenotemark{c,d,i,3,4}   &0.094  &PC/{\it F814W}\tablenotemark{$\ast$}      &-22.14$^{\pm 0.08}$       &2.80$^{\pm 0.10}$       &18.66$^{\pm 0.13}$\\
PG 1444+407      &14 46 45  &+40 35 06  &0.267   &12.43\tablenotemark{c,g}         &0.014  &WF3/{\it F606W}\tablenotemark{$\star$}    &-24.24                  &4.97                  &17.80\\
HB89 1549+203    &15 52 02  &+20 14 02  &0.250   &12.33\tablenotemark{b,c,d,3,4}   &0.054  &WF2/{\it F675W}\tablenotemark{$\ddagger$} &-22.49                  &3.48                  &18.78\\
HB89 1635+119    &16 37 46  &+11 49 49  &0.146   &11.83\tablenotemark{b,c,d}       &0.052  &WF2/{\it F675W}\tablenotemark{$\ast$}     &-23.18$^{\pm 0.02}$       &4.16$^{\pm 0.03}$       &18.47$^{\pm 0.06}$\\
HB89 1821+643    &18 21 57  &+64 20 36  &0.297   &13.05\tablenotemark{c,j}         &0.043  &WF3/{\it F791W}\tablenotemark{$\dagger$}  &-24.81$^{\pm 0.02}$       &12.52$^{\pm 0.13}$      &19.23$^{\pm 0.05}$\\
HB89 2215-037    &22 17 47  &-03 32 38  &0.242   &12.37\tablenotemark{b,3,4}       &0.106  &PC/{\it F702W}\tablenotemark{$\ast$}      &-23.88$^{\pm 0.02}$       &6.79$^{\pm 0.06}$       &18.84$^{\pm 0.08}$\\
\enddata
\tablecomments{{\footnotesize
PC $=$ Planetary Camera; WF2 $=$ Wide Field 2; WF3 $=$ Wide Field 3; WF4 $=$ Wide Field 4.   
All rest-frame {\it I}-band photometry is in Vega magnitudes and has been corrected for Galactic Reddening.
The {\it A}$_{\lambda}$ scaling factors used were:  
{\it A}$_{\rm F606W}$ $=$ 2.41;
{\it A}$_{\rm F675W}$ $=$ 2.52;
{\it A}$_{\rm F702W}$ $=$ 1.94;
{\it A}$_{\rm F791W}$ $=$ 1.74;
{\it A}$_{\rm F814W}$ $=$ 1.54.
The {\it R}$_{\rm eff}$ listed in this table are {\it equivalent} radii. 
(a) Galactic reddening values from \citep{1998ApJ...500..525S};
(b) Fluxes from NASA$/$IPAC Scan Processing and Integration tool;
(c) 12$\micron$ flux from {\it WISE};
(d) 22$\micron$ flux from {\it WISE};
(e) computed only from 12 and 22$\micron$ {\it WISE} fluxes;
(f) \cite{1988AJ.....95..307I};
(g) \cite{1989ApJ...347...29S}; 
(h) \cite{2000A&A...354..453H} based on ISO photometry;
(i) \cite{2003A&A...402...87H} based on ISO photometry;
(j) IRAS Faint Source Catalog; 
(1) upper limits (3$\times${\it r.m.s.}) used for {\it f}$_{\rm 12}$; 
(2) upper limits (3$\times${\it r.m.s.}) used for {\it f}$_{\rm 25}$; 
(3) upper limits (3$\times${\it r.m.s.}) used for {\it f}$_{\rm 60}$;
(4) upper limits (3$\times${\it r.m.s.}) used for {\it f}$_{\rm 100}$.
The IRAS fluxes for 3C59 were computed from 3$\sigma$ upper limits only and 
noted in parentheses above.
($\dagger$) {\it HST/WFPC2} Photometry from \cite[][Floyd, {\it private communication}]{2004MNRAS.355..196F};
($\ast$) {\it HST/WFPC2} Photometry from \cite[][Hamilton, {\it private communication}]{2002ApJ...576...61H,2008ApJ...678...22H};
($\ddagger$) {\it HST/WFPC2} Photometry from \cite{2003MNRAS.340.1095D}. Although {\it R}-band values are listed in the paper, they are actually values for the
{\it F675W} filter and were not transformed to Cousins {\it R}-band;
($\star$) {\it HST/WFPC2} Photometry from \cite{1997ApJ...479..642B}.}}
\end{deluxetable}

\clearpage
\begin{deluxetable}{lcc}
\tabletypesize{\footnotesize}
\tablewidth{0pt}
\tablenum{2}
\pagestyle{empty}
\tablecaption{QSO Host Kinematic Properties}
\tablecolumns{3}
\tablehead{
\colhead{Galaxy} &
\colhead{Optical $\sigma$$_{\circ}$} &
\colhead{{\it M}$_{\rm dyn}$} \\
\colhead{Name} &
\colhead{(km s$^{-1}$)} &
\colhead{($\times$10$^{11}${\it M}$_{\odot}$)}
}
\startdata
\cutinhead{Radio Loud QSOs}
PKS 0736+01                      &348 $^{\pm 83}$   &14.28$^{\pm 7.62}$    \\
PG 1226+023                      &339 $^{\pm 58}$   &13.54$^{\pm 5.15}$    \\
PG 1302-102                      &386 $^{\pm 72}$   &17.13$^{\pm 7.13}$    \\
PKS 2135-14                      &310 $^{\pm 106}$  &13.78$^{\pm 10.50}$   \\
4C +31.63\tablenotemark{a}       &318 $^{\pm 49}$   &9.92$^{\pm 3.39}$     \\
PG 2349-014\tablenotemark{a}    &279 $^{\pm 64}$   &19.78$^{\pm 10.09}$   \\
\cutinhead{Radio Quiet QSOs}
HB89 0054+144                    &167 $^{\pm 11}$   &1.76$^{\pm 0.26}$     \\
PG 1444+407                      &313 $^{\pm 22}$   &6.81$^{\pm 1.07}$     \\
\enddata
\tablecomments{{\footnotesize All $\sigma$$_{\circ}$ values are measured using the
\ion{Ca}{II} H\&K stellar absorption line over the 0.385-0.42 $\micron$ 
wavelength range.  All values of $\sigma$$_{\circ}$ have been corrected to a 
common aperture of 1.53 kpc.   The average radii of the extracted apertures
are listed in Table 3 of \cite{2008AJ....136.1587W}.
(a) The average radius is used for multiple extracted positions \citep{2008AJ....136.1587W}.}}
\end{deluxetable}


{\vspace{-2in}
\begin{figure}
\section{Images, Light Profiles, \& Spectra of the ULIRG Sample}
{
\plotone{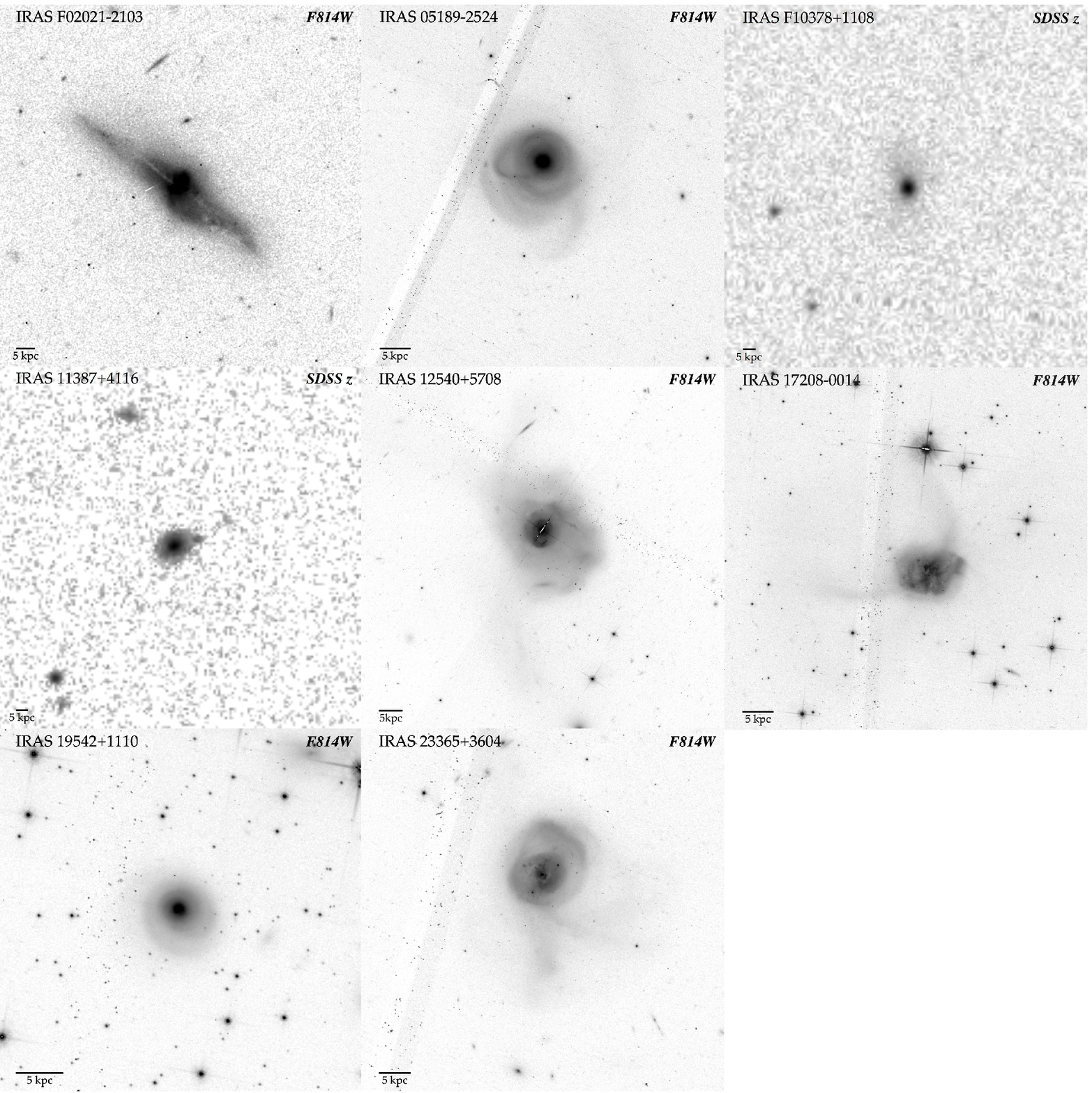}}
\caption{HST {\it F814W} or SDSS {\it z} images for the 8 advanced ULIRGs. 
The images are presented in reverse grayscale with a logarithmic stretch. 
Overplotted on each image is a horizontal solid bar representing 5 kpc. In each
image North is up.}
\end{figure}}

\begin{figure}
\epsscale{1.1}
\plotone{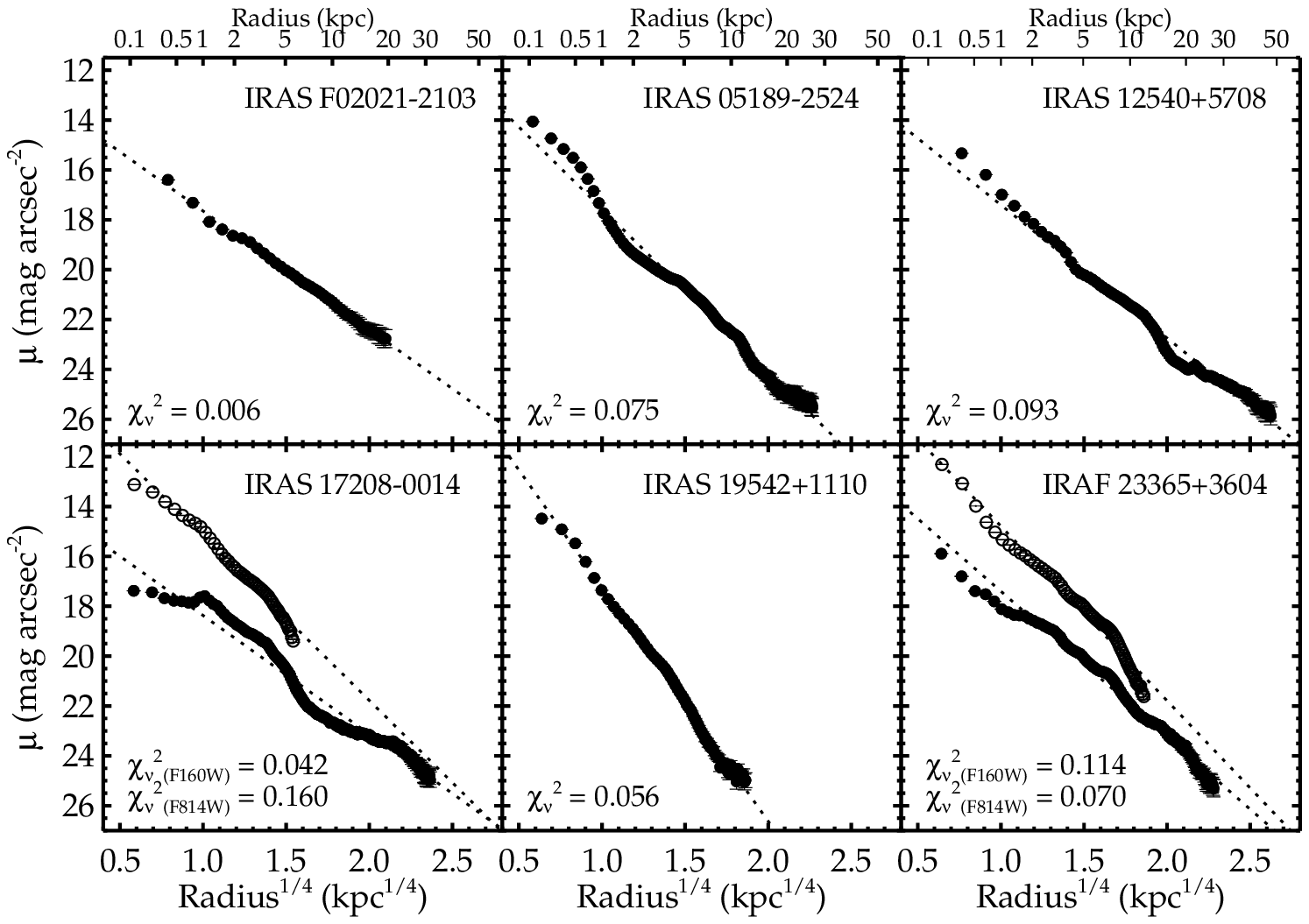}
\caption{Plotted here are the surface brightness profiles in the {\it F814W}-band 
(filled circles) for the 6 ULIRGs observed with {\it HST} using ACS/WFC or WFPC2.  
Also shown are profiles for two ULIRGs observed with {\it HST} using the NIC2 camera at 
{\it F160W} (open circles).  1$\sigma$ standard errors are over-plotted on each point.  
The surface brightness profiles are measured out to a S/N$=$3 over the background.  
All profiles are measured using circular apertures.  The plotted points are equally 
spaced and linear, corresponding to 3 pixels for ACS, 2 pixels for WFPC2, and 2 pixels 
for NIC2 (a radius of 9 pixels was used for IRAS 12540+5708 to compensate for the 
saturated/masked central region).  The light dashed line in each plot 
represents the best-fit de Vaucouleurs {\it r}$^{1/4}$ fit to all of the data. 
The $\chi$$_{\nu}$$^{2}$ of the best fit is shown in each panel.  The two profiles
plotted in the panels for IRAS 17208-0014 and IRAS 23365+3604 have not been shifted.  
They represent the actual values.  These two ULIRGs also show significantly redder 
$(I-H)$ colors at {\it R} $<$ 1 kpc than the value of 1.77-1.79 expected from a typical
elliptical galaxy \citep{1978ApJ...220...75F,1999ApJS..124..127P}, consistent with
the results from Figure 13 of Paper I.
}
\end{figure}

\begin{figure}
\epsscale{1.25}
\plotone{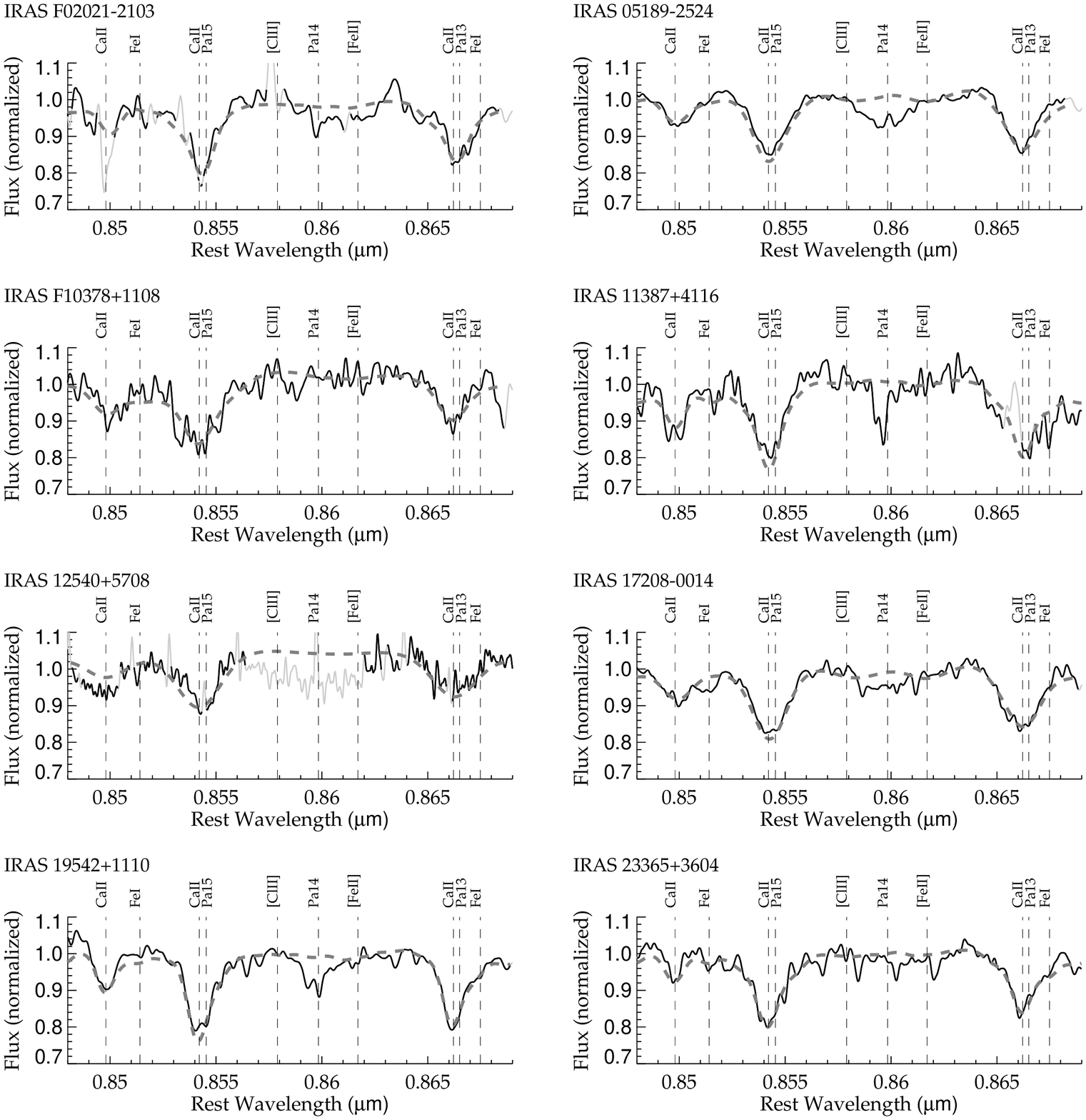}
\caption{Shown here are the CaT spectra of the ULIRGs observed with ESI on Keck-2.  The solid black lines show
the actual spectrum, the light grey lines show masked bad pixels and/or emission lines.  
The thick dashed line shows the best-fit convolved template.  Also shown are the positions of known emission and 
absorption lines within the wavelength range (Note: this does not mean that all of these lines are detected.)}
\end{figure}
\clearpage

\begin{deluxetable}{lcccccccc}
\tabletypesize{\footnotesize}
\tablewidth{0pt}
\tablenum{1}
\pagestyle{empty}
\tablecaption{Summary of KS Tests:
Can the Null Hypothesis be Rejected at the 95$\%$ level?}
\tablecolumns{9}
\tablehead{
\colhead{Comparison} &
\colhead{Fundamental} &
\colhead{$\sigma$$_{\circ}$} &
\colhead{{\it M}$_{\rm Dyn}$-{\it M}/{\it L}} &
\colhead{{\it M}$_{\rm Dyn}$} &
\colhead{Kormendy\tablenotemark{a}} &
\colhead{Log {\it R}$_{\rm eff}$\tablenotemark{a}} &
\colhead{$<$$\mu$$_{\rm I}$$>$$_{\rm eff}$\tablenotemark{a}} &
\colhead{Log {\it L}$_{\rm IR}$}\\
\colhead{Sample} &
\colhead{Plane (2D)} &
\colhead{}  & 
\colhead{(2D)} &
\colhead{} &
\colhead{Relation (2D)} &
\colhead{} &
\colhead{} &
\colhead{({\it L}$_{\odot}$)}
}
\startdata
RLQs              &No (89$\%$)            &Yes (26$\%$)          &Yes\tablenotemark{d} (72$\%$)   &Yes\tablenotemark{e} (87$\%$)   &No                    &No\tablenotemark{h} &No   &No\\
RQQs              &\nodata                &\nodata               &\nodata                         &\nodata                         &No                    &No                  &No   &No\\
All QSOs          &No (100$\%$)           &No  (97$\%$)          &Yes  (27$\%$)                   &No  (100$\%$)                   &No                    &No                  &No   &No\\
SDSS Ellipticals  &Yes\tablenotemark{b}   &Yes\tablenotemark{c}  &Yes\tablenotemark{b}            &Yes\tablenotemark{f}            &No\tablenotemark{g}   &No\tablenotemark{i} &No   &\nodata\\
\enddata
\tablecomments{{\footnotesize This table presents a summary of the 1D and 2D KS tests performed between the distributions
of the ULIRGs and the various samples listed in Column 1.
The term ``2D'' refers to the two-dimensional KS test.  Otherwise the KS test results
presented here are for the standard two-sided KS test which compares the distributions of two
empirical samples.  Due to the large errors associated with the RLQ $\sigma$$_{\circ}$ values,
the KS tests also have been run using Monte Carlo simulations in which each RLQ $\sigma$$_{\circ}$
is randomly assigned a value of $\sigma$$_{\circ}$$\pm$$\Delta$$\sigma$$_{\circ}$.
Test results above which include the RLQ $\sigma$$_{\circ}$ parameter show a parentheses
indicating what percent of 10,000 Monte Carlo simulations the result occurs. 
Although IRAS 19542+1110 is 3.9$\sigma$ outlier from the FP, excluding it from the KS tests in columns 2-5
does not change the results presented above.
(a) Comparison with larger photometric sample of QSOs (28 RLQs and 25 RQQs);
(b) No for 165-225 km s$^{-1}$ and 225-420 km s$^{-1}$ bins;
(c) The Null Hypothesis can also be rejected at the 99$\%$ level;
(d) The Null Hypothesis can also be rejected at the 99$\%$ level but only 13$\%$ of the time;
(e) The Null Hypothesis can also be rejected at the 99$\%$ level but only 12$\%$ of the time;
(f) No for the  225-420 km s$^{-1}$ bins;
(g) The Null Hypothesis can be rejected when IRAS 19542+1110 is excluded.  The Null Hypothesis
cannot be rejected for the 165-225 km s$^{-1}$ and 225-420 km s$^{-1}$ bins whether or not
IRAS 19542+1110 is excluded;
(h) If only the 6 RLQs in the kinematic sub-sample are compared with the ULIRGs, then the Null Hypothesis can be rejected;
(i) When IRAS 19542+1110 is excluded the Null Hypothesis can be rejected.}}
\end{deluxetable}

\clearpage

\section{Cosmic Ray Rejection Algorithm and Bad Pixel Mask}
\indent Most {\it HST}/ACS programs employ either {\tt CR-SPLIT}, (two separate exposures 
at the same pointing), or multiple (e.g. $>$ 2) dithered positions to remove cosmic rays 
(CRs) and artifacts.   The {\tt CR-SPLIT} mode is most appropriate for programs where
the absence of data in the gap will not impact the science (small targets or point 
sources).  {\tt MULTIDRIZZLE} compares the two images and flags pixels that have 
changed significantly between the two exposures.  Dithering will ``fill in'' the chip gap 
and allow for the recovery of information in the chip gap.  With at least 3 dither 
positions the same technique for removing CRs in the chip gap can be used.
Program 10592 used a two position dithering scheme ({\tt ACS-WFC-DITHER-LINE}) 
that shifts the image 5 pixels in X and 60 pixels in Y.  This fills the chip gap but
with data from one exposure only (each gap is filled by information from the other
exposure).  Thus, the final images contained significant CR hits and hot pixels in the 
chip gap. Because the ULIRGs were not centered on either chip, but 
centered in the {\it ACS/WFC} FOV (aperture {\tt WFC}) the chip gap runs through the outer 
regions of the galaxy, impacting the science data.  Because more objects were affected
by this than in Paper I, an improved and more automated algorithm was developed to 
create a bad pixel mask.  First, a zero level background 
image was created by identifying the median background flux levels of the multidrizzled 
final image and replacing pixels at or below these values with a value of zero (using 
{\tt IMREPLACE}).  Second, the zero-level background image was passed through a median 
filter using a 15$\times$15 pixel filter box, creating a new filtered image.  
Third, the zero-level background image was divided by the median filtered image.  In this
divided image, all pixels with flux values above the maximum pixel value in the nucleus 
were set to a value of 1 (bad), the remaining pixels were to set to a value of 0 (good).  
This pixel mask proved successful for identifying saturated stars, diffraction spikes, 
and elongated CRs in the gap and areas covered by only one pointing.  This pixel mask 
was then combined with one created from pixels flagged bad by {\tt MULTIDRIZZLE}, 
and a pixel mask created from the positions of foreground stars and background
galaxies.  \\
\\


\end{document}